\documentclass[twocolumn,twocolappendix]{aastex631}

\usepackage{amsmath}
\usepackage{empheq}
\usepackage{mathrsfs}
\usepackage{textcomp}
\usepackage{enumitem}   
\usepackage{gensymb}
\usepackage{hyperref}
\usepackage{graphicx}
\usepackage[caption=false]{subfig}
\usepackage{multirow}
\usepackage{longtable}
\usepackage[flushleft]{threeparttable}
\usepackage{booktabs} % To thicken table lines
\usepackage{CJK}
\hypersetup{colorlinks, linkcolor={blue}, citecolor={blue}, urlcolor={blue}}

\definecolor{DarkOrange}{RGB}{204, 85, 0}
\definecolor{LincolnGreen}{RGB}{17, 102, 0}
\def\ion#1#2{#1$\;${\footnotesize\rm{#2}}\relax}

\usepackage{xspace}

\newcommand\nustar{\textit{NuSTAR}\xspace}
\newcommand\nicer{\textit{NICER}\xspace}
\newcommand\chandra{\textit{Chandra}\xspace}
\newcommand\rosat{\textit{ROSAT}\xspace}
\newcommand\swift{\textit{Swift}\xspace}
\newcommand\maxi{\textit{MAXI}\xspace}
\newcommand\srg{\textit{SRG}\xspace}
\newcommand\rxte{\textit{RXTE}\xspace}
\newcommand\integral{\textit{INTEGRAL}\xspace}
\newcommand\target{AT2019wey\xspace}
\newcommand\cmcm{$\rm cm^{-2}$\xspace}
\newcommand\cps{$\rm count\,s^{-1}$\xspace}

\def \caltech {{Cahill Center for Astronomy and Astrophysics, 
	California Institute of Technology, Pasadena, CA 91125, USA}}
\def \mit {{MIT Kavli Institute for Astrophysics and Space Research, 
    70 Vassar Street, Cambridge, MA 02139, USA}}

\def \nrl {{Space Science Division, 
            U.S. Naval Research Laboratory, 
            Washington, DC 20375, USA}}
            
\def \gsfc {{Astrophysics Science Division, NASA Goddard Space Flight Center, Greenbelt, MD 20771, USA}}

\shorttitle{AT2019wey: X-ray Observations}
\begin{document}

\title{A Comprehensive X-ray Report on AT2019wey}

\correspondingauthor{Yuhan Yao}
\email{yyao@astro.caltech.edu}

\author[0000-0001-6747-8509]{Yuhan Yao}
\affiliation{\caltech}

%%% PIs
\author[0000-0001-5390-8563]{S. R. Kulkarni}
\affiliation{\caltech}

\author{K. C. Gendreau}
\affiliation{Mail Code 662.1, Goddard Space Flight Center, Greenbelt, MD 20771, USA}

%%% Spectral Analysis
\author[0000-0002-6789-2723]{Gaurava K. Jaisawal}
\affiliation{National Space Institute, Technical University of Denmark, Elektrovej 327-328, DK-2800 Lyngby, Denmark}

\author[0000-0003-1244-3100]{Teruaki Enoto} % confirmed
\affiliation{Extreme Natural Phenomena RIKEN Hakubi Research Team, 
    Cluster for Pioneering Research, RIKEN, \\
    2-1 Hirosawa, Wako, Saitama 351-0198, Japan}

\author[0000-0002-1984-2932]{Brian W. Grefenstette}
\affiliation{\caltech}

\author[0000-0002-6492-1293]{Herman L. Marshall}
\affiliation{\mit}

\author[0000-0003-3828-2448]{Javier~A.~Garc\'ia}
\affiliation{\caltech}
\affiliation{Dr. Karl Remeis-Observatory and Erlangen Centre for Astroparticle Physics, Sternwartstr.~7, 96049 Bamberg, Germany}

\author[0000-0002-8961-939X]{R.~M.~Ludlam}\thanks{NASA Einstein Fellow}
\affiliation{\caltech}

%%% Timing Analysis g
\author[0000-0002-8403-0041]{Sean N. Pike}
\affiliation{\caltech}

\author[0000-0002-0940-6563]{Mason Ng}
\affiliation{\mit}

\author[0000-0003-4498-9925]{Liang Zhang}
\affiliation{Department of Physics and Astronomy, 
    University of Southampton, Southampton, SO17 1BJ, UK}

\author[0000-0002-3422-0074]{Diego Altamirano}
\affiliation{Department of Physics and Astronomy, 
    University of Southampton, Southampton, SO17 1BJ, UK}
    
\author[0000-0002-3850-6651]{Amruta Jaodand}
\affiliation{\caltech}

\author[0000-0003-1673-970X]{S. Bradley Cenko}
\affiliation{\gsfc}
    
\author[0000-0003-4815-0481]{Ronald A. Remillard}
\affiliation{\mit}

\author{James F. Steiner}
\affiliation{Harvard-Smithsonian Center for Astrophysics, 
    60 Garden St., Cambridge, MA 02138, USA}

%%% MAXI team
\author[0000-0003-0939-1178]{Hitoshi Negoro}
\affiliation{Department of Physics, Nihon University, 1-8,
    Kanda-Surugadai, Chiyoda-ku, Tokyo 101-8308}
    
%%% Others with contribution
\author{Murray Brightman}
\affiliation{\caltech}

\author[0000-0002-7851-9756]{Amy Lien}
\affiliation{Center for Research and Exploration in Space Science and Technology (CRESST) and NASA Goddard Space Flight Center, Greenbelt, MD 20771, USA}
\affiliation{Department of Physics, University of Maryland, Baltimore County, 1000 Hilltop Circle, Baltimore, MD 21250, USA}
    
\author[0000-0002-4013-5650]{Michael T. Wolff}
\affiliation{\nrl}

\author[0000-0002-5297-5278]{Paul S. Ray}
\affiliation{\nrl}

\author{Koji Mukai}
\affiliation{CRESST II and X-ray Astrophysics Laboratory, NASA/GSFC, Greenbelt, MD 20771, USA}
\affiliation{Department of Physics, University of Maryland, Baltimore County, 1000 Hilltop Circle, Baltimore, MD 21250, USA}

\author[0000-0002-9249-0515]{Zorawar Wadiasingh}
\affiliation{\gsfc}
\affiliation{Universities Space Research Association (USRA), Columbia, MD 21046, USA}

\author{Zaven Arzoumanian}
\affiliation{\gsfc}

\author{Nobuyki Kawai}
\affiliation{Department of Physics, Tokyo Institute of Technology, 2-12-1 Ookayama, Meguro-ku, Tokyo 152-8551}

\author{Tatehiro Mihara}
\affiliation{Cosmic Radiation Laboratory, RIKEN, 2-1 Hirosawa, Wako, Saitama 351–198}

\author[0000-0001-7681-5845]{Tod E. Strohmayer}
\affiliation{\gsfc}

\begin{abstract}
Here, we present \maxi, \swift, \nicer, \nustar and \chandra observations of the X-ray transient AT2019wey (SRGA\,J043520.9$+$552226, SRGE\,J043523.3$+$552234). From spectral and timing analyses we classify it as a Galactic low-mass X-ray binary (LMXB) with a black hole (BH) or neutron star (NS) accretor.
\target stayed in the low/hard state (LHS) from 2019 December to 2020 August 21, and the hard-intermediate state (HIMS) from 2020 August 21 to 2020 November.
For the first six months of the LHS, \target had a flux of $\sim 1$\,mCrab, and displayed a power-law X-ray spectrum with photon index $\Gamma = 1.8$.
From 2020 June to August, it brightened to $\sim 20$\,mCrab. 
Spectral features characteristic of relativistic reflection became prominent.
On 2020 August 21, the source left the ``hard line'' on the rms--intensity diagram, and transitioned from LHS to HIMS. The thermal disk component became comparable to the power-law component.
A low-frequency quasi-periodic oscillation (QPO) was observed. 
The QPO central frequency increased as the spectrum softened.
No evidence of pulsation was detected.
We are not able to decisively determine the nature of the accretor (BH or NS). 
However, the BH option is favored by the position of this source on the $\Gamma$--$L_{\rm X}$, $L_{\rm radio}$--$L_{\rm X}$, and $L_{\rm opt}$--$L_{\rm X}$ diagrams.
We find the BH candidate XTE\,J1752$-$223 to be an analog of AT2019wey. 
Both systems display outbursts with long plateau phases in the hard states. 
We conclude by noting the potential of \srg in finding new members of this emerging class of low luminosity and long-duration LMXB outbursts.
\end{abstract}

%\keywords{Accretion (14); Black holes (162); Low-mass x-ray binary stars (939); X-ray transient sources (1852); Transient sources (1851); X-ray sources (1822); X-ray binary stars (1811)}

\section{Introduction}
%\subsection{Low-mass X-ray Binaries and X-ray States}
% begin with an introduction to LMXBs, 
Low-mass X-ray binaries (LMXBs) contain a neutron star (NS) or black hole (BH) accretor and a low-mass ($\lesssim 2\,M_\odot$) companion star. To first order, LMXBs with higher mass transfer rates ($\dot M$) can keep the accretion disks fully ionized, and are observed as persistent sources. Systems with lower $\dot M$ exhibit prolific outbursts, which are popularly attributed to the thermal-viscous instability \citep{Tanaka1996, Done2007, Coriat2012}. 

Transient BH LMXBs have been observed in several distinct X-ray states, including the steep power-law (SPL) state (also known as the very high state), the high/soft state (HSS; also known as the thermal state), the intermediate state (IMS), the low/hard state (LHS), and the quiescent state \citep{Fender2004, Remillard2006, McClintock2006, Belloni2010, Gilfanov2010, Zhang2013}. This classification scheme relies primarily on the shape of the 1--20\,keV energy spectrum, the spectral hardness (defined as the ratio of count rates in the hard and soft energy bands), the fractional rms variability integrated over a range of frequencies, and the presence of quasi-periodic oscillation (QPO) in the power density spectrum (PDS). 

% Evolution
The evolution of a BH outburst is traditionally described as a counterclockwise q-shaped track in the hardness--intensity diagram (HID; \citealt{Homan2005}), following the sequence of quiescence $\rightarrow$ LHS $\rightarrow$ IMS  $\rightarrow$ HSS $\rightarrow$ IMS $\rightarrow$ LHS $\rightarrow$ quiescence.
% LHS
In the LHS, the X-ray spectrum is dominated by a non-thermal power-law component with a photon index ($\Gamma$) of of 1.5--2.0. This state is commonly accompanied by strong aperiodic variability and low-frequency QPO (LFQPO; 0.1--30\,Hz), with fractional rms of $\sim30$\%. 
% IMS
In the IMS, a thermal disk component with a color temperature of 0.1--1\,keV appears and $\Gamma$ softens to 2.0--2.5. The IMS can be further separated into the hard-intermediate state (HIMS), where the fractional rms is $\sim10$--20\%, and the soft-intermediate state (SIMS), where the fractional rms is a few percent \citep{Belloni2010}. 
% HSS
In the HSS, the thermal accretion disk becomes the dominant component in the X-ray spectrum. Meanwhile, QPOs are absent or very weak, and the fractional rms drops to $\sim1$\%. 
% SPL
Occasionally, the outburst goes into the SPL state as it approaches the Eddington luminosity $L_{\rm Edd}$. Here, the spectrum is dominated by a power-law spectral component with a photon index of $\Gamma \sim 2.5$. 

Although a good number of BH LMXB outbursts follow the above hysteresis loop of state transition, some remain in the LHS throughout the entire outburst \citep{Belloni2002XTEJ1550, Brocksopp2004, Sidoli2011}, and some only transition between the LHS and the HIMS \citep{Ferrigno2012, Soleri2013, Capitanio2009}. By analyzing the BH outbursts between 1996 and 2015, \citet{Tetarenko2016} show that $\sim40$\% of them only stay in the LHS or HIMS. These ``hard-only'' outbursts (also termed as failed outbursts) are generally associated with lower peak luminosities.

NS LMXBs can be broadly classified into Z sources and atoll sources, named after the tracks they trace out in the HID and X-ray color-color diagram \citep{Hasinger1989, vanderKlis2006}. Z sources are generally bright ($L_{\rm X}\gtrsim 0.5L_{\rm Edd}$). Atoll sources are further divided into bright atoll sources (BA; $L_{\rm X}\sim0.3$--$0.5L_{\rm Edd}$) and ordinary atoll sources ($L_{\rm X}\sim 0.01$--$0.3L_{\rm Edd}$). The X-ray spectra of Z and BA sources remain very soft, while ordinary atolls mostly follow the hysteresis pattern of state transition observed in BH LMXBs \citep{MunozDarias2014}. 

BHs can be identified via dynamical measurements of the mass of the compact object. NS LMXBs can be selected by the existence of thermonuclear X-ray bursts from nuclear burning of the accreted material on the NS surface, or coherent pulsations caused by the magnetic field and NS rotation \citep{Lewin1993, Done2007, Bhattacharyya2009}. 

Kilo-hertz QPOs have only been observed in NS systems \citep{vanDoesburgh2018}. Furthermore, in the hard state, NSs have systematically lower values of Compton $y$-parameter and electron temperature \citep{Banerjee2020}, as well as higher values of $\Gamma$ \citep{Wijnands2015}. Compared with BH LMXBs, the PDS of NS systems can show broad-band noise at frequencies up to $500$\,Hz \citep{Sunyaev2000}.
In the soft state, the spectra of NS LMXBs are harder than those of BH systems due to an additional thermal emission from the boundary layer with a blackbody temperature of $\sim2.4$\,keV \citep{Done2003, Gilfanov2003}. 

\subsection{AT2019wey}
AT2019wey was discovered as an optical transient by the ATLAS optical survey in 2019 December \citep{Tonry2019}. 
It rose to prominence in 2020 March with the discovery of X-ray emission by the eROSITA \citep{Predehl2021} and the Mikhail Pavlinsky ART-XC \citep{Pavlinsky2021} telescopes on board the \textit{Spektrum-Roentgen-Gamma} (\srg) satellite \citep{Sunyaev2021}. Upon detection, the X-ray flux was 0.36\,mCrab in the 0.3--8\,keV band and 0.6\,mCrab in the 4--12\,keV band \citep{Mereminskiy2020}. We note that there is no point source detected at the position of \target in the 2nd \rosat All-Sky Survey Point Source Catalog (2RXS; \citealt{Boller2016}), providing a historical upper limit of $\sim10\,\mu$Crab in 0.1--2.4\,keV. 

Initially \target was thought to be a supernova \citep{Mereminskiy2020} and subsequently proposed to be a BL Lac object \citep{Lyapin2020}. \citet{Yao2020_13932} reported the detection of hydrogen lines at redshift $z=0$, and proposed \target to be a Galactic accreting binary. 

Here, we report comprehensive X-ray observations from the beginning of 2019 January to the end of 2020 November. We find that AT2019wey is consistent with the spectral and timing behavior expected from a LMXB harbouring a BH or NS accretor. Elsewhere we report  the multi-wavelength observations of this source \citep[][hereafter Paper II]{Yao2020}.

This paper is organized as follows. In Section~\ref{sec:obs} we describe the X-ray observations and data reduction. We present the analysis of light curves in Section~\ref{sec:result}, including the evolution of hardness (Section~\ref{subsec:colors}) and the timing properties (Section~\ref{subsec:timing}). The spectral analysis including multi-mission joint analysis can be found in Section~\ref{sec:spectral}. In Section~\ref{sec:discussion} we present the inferences from the X-ray analysis. We are not able to decisively identify the nature of the accretor (BH or NS). However, we find the best analog to AT2019wey is a candidate BH LMXB system. We conclude in Section~\ref{sec:conclusion}.\footnote{UT time is used throughout the paper.}

\section{Observations and Data Reduction} \label{sec:obs}

The data shown here were obtained using the Neutron Star Interior Composition Explorer (\nicer; \citealt{Gendreau2016}), the Nuclear Spectroscopic Telescope ARray (\nustar; \citealt{Harrison2013}), the \chandra X-ray Observatory \citep{Chandra2020}, the \textit{Neil Gehrels Swift Observatory} \citep{Gehrels2004}, and the Monitor of All-sky X-ray Image (\maxi) mission \citep{Matsuoka2009}. As of the time of submission of the paper (2020 November), the source was still active.

\subsection{\nicer} \label{subsec:obs_nicer}

AT2019wey was observed by the X-ray Timing Instrument (XTI) on board \nicer over the period from 2020 August 4 to 2020 September 30 (PI: K.~Gendreau). The observation log is shown in Table~\ref{tab:nicer}. \nicer is a soft X-ray telescope on board the International Space Station (ISS). \nicer is comprised of 56 co-aligned concentrator X-ray optics, each paired with a single-pixel silicon drift detector. Presently, 52 detectors are active with a net peak effective area of $\sim1900~{\rm cm}^2$ at 1.5\,keV, and 50 of these were selected (excluding modules 14 and 34) to make the light curves reported in this paper. The \nicer observations were processed using \texttt{HEASoft} version 6.27 and the \nicer Data Analysis Software (\texttt{nicerdas}) version 7.0 (2020-04-23\_V007a). 

To generate a background-subtracted light curve, we first defined good time intervals (GTIs) with as much data as possible. Then we computed the background using the \texttt{nibackgen3C50} tool \citep{Remillard2021}. For each GTI, we explicitly subtracted the background-predicted spectrum from the raw extraction to get the source net spectrum. We also removed GTIs with $|hbg|$\footnote{$hbg$ is the count rate in the 13--15\,keV band, which is beyond the effective area of the concentrator optics, as defined in \citet{Remillard2021}.}$> 0.07$, to exclude GTIs with less accurate background subtraction. Finally, we computed count rate in five energy bands: 0.4--1.0\,keV, 1--2\,keV, 2--4\,keV, 4--12\,keV, and 0.4--12\,keV. 

To generate spectral files, we used \texttt{nimaketime} to select GTIs that occurred when the particle background was low (KP$<$5 and COR\_SAX$>$4). We removed times of extreme optical light loading and low Sun angle by selecting FPM\_UNDERONLY\_COUNT$<$200 and SUN\_ANGLE$>$60. Using \texttt{niextract-events}, the GTIs were applied to the data selecting EVENT\_FLAGS=bxxx1x000 and PI energy channels between 25 and 1200, inclusive. For more information on the \nicer screening flags, see \citet{Bogdanov2019}. The resulting event files were loaded into \texttt{xselect} to extract a combined spectrum after filtering. Systematic errors of 1\% in the 2--10\,keV band and 5\% in the 0.8--2\,keV band were added via \texttt{grppha}. A background spectrum was generated using the \texttt{nibackgen3C50} tool for each cleaned event file and ufa event file pair. These were then combined into a single background spectrum that was weighted by the duration of each observation. Each spectrum was grouped into channels by considering a minimum of 32 counts per channel bin.

\subsection{\nustar}

We obtained three epochs of Target of Opportunity (ToO) observations using the hard X-ray telescope \nustar (PI: Y.~Yao, Table~\ref{tab:nustar}). In this paper, we report the analysis for the first two sequences (sequence IDs 90601315002 and 90601315004, hereafter 002 and 004, respectively). The third sequence was carried out jointly with the Hard X-ray Modulation Telescope (HXMT; \citealt{ZhangSN2020}), and will be reported in Tao et al. in prep.

The focal plane of \nustar consists of two photon counting detector modules (FPMA and FPMB). The data were processed using the \nustar Data Analysis Software (\texttt{nustardas}) v2.0.0 along with the 2020423 \nustar CALDB using the default data processing parameters. Cleaned event files were produced with \texttt{nupipeline}. The event files were then barycentered and corrected for clock offset using \texttt{barycorr}\footnote{We used the clockfile version \texttt{v108}. See \citet{Bachetti2021} for a description of the clockfile generation.}. 

To generate \nustar light curves, we filtered the events using source regions with radii of $60^{\prime\prime}$ and $90^{\prime\prime}$ for 002 and 004, respectively. We chose to use a larger source region for observation 004 due to its higher count rate. We only retained events with a photon energy between 3 and 78\,keV. For each observation, we were thus left with two lists of filtered and barycenter-corrected events --- one for FPMA and one for FPMB. Using \texttt{Stingray} \citep{Huppenkothen2019}, we produced light curves for each of these event lists. We binned the light curves with a time resolution of $\approx2$\,ms. %$512^{-1}$\,s. 
\texttt{Stingray} automatically applied the GTIs recorded by the instrument.

To generate the spectra for FPMA and FPMB, source photons were extracted from a circular region with a radius of 60$^{\prime\prime}$ centered on the apparent position of the source in both FPMA and FPMB. For 002 the background was extracted from a 100$^{\prime\prime}$ region located on the same detector; For 004 the source was bright enough that a smaller portion of the field-of-view could be used to estimate the background, so the background was extracted from a 60$^{\prime\prime}$ region.

\subsection{\chandra}
We requested and were granted 25\,ks of \chandra director’s discretionary time (PI: S.~R.~Kulkarni; ${\rm OBSID}=24651$) to obtain a high-energy transmission grating spectrometer (HETGS; \citealt{Markert1994, Canizares2005}) spectrum using the Advanced CCD Imaging Spectrometer (ACIS; \citealt{Garmire2003}). The HETGS is composed of two sets of gratings (see, e.g., Chapter 2 of \citealt{Chandra2020}): the medium-energy gratings (MEGs) cover the 0.4--7\,keV energy band, and the high-energy gratings (HEGs) cover the 0.8--10.0\,keV band. The observation was carried out in the timed event (TE) mode around the maximum soft X-ray luminosity of \target. During the exposure (from 2020-09-20T17:43 to 2020-09-21T01:12), the source count rate varied between 23.1\,\cps to 24.5\,\cps. 

To generate spectral files for the source and the background, we extracted the plus and minus first-order ($m = \pm 1$) MEG and HEG data from the $-1$ and the $+1$ arms of the MEG and HEG gratings. We used the \texttt{CIAO} tool \texttt{tgextract}. \texttt{CIAO} version 4.12.1 and the associated CALDB version 4.9.3 were used in the data reduction.
Spectral redistribution matrix files and effective area files were generated with \texttt{mkgrmf} and \texttt{mkgarf}.

\begin{figure*}[htbp!]
	\centering
	\includegraphics[width=\textwidth]{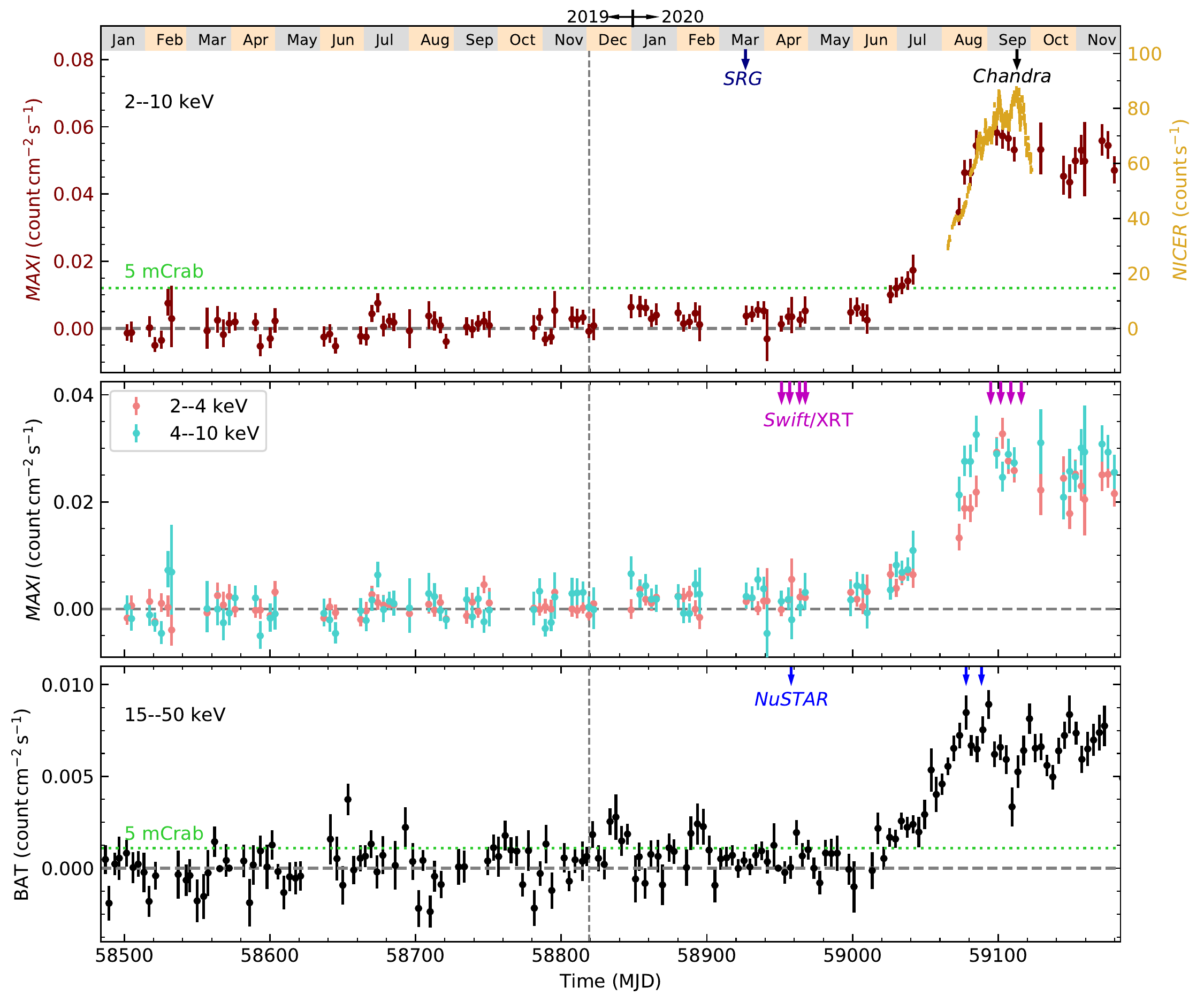}
	\caption{\textit{Upper}: \nicer and \maxi light curves of \target in the 2--10\,keV band. The dark blue arrow marks the epoch of \srg discovery. The black arrow marks the epoch of \chandra observation.  \textit{Middle}: \maxi light curves in the 2--4\,keV and 4--10\,keV bands. The magenta arrows mark epochs of \swift/XRT observations (Table~\ref{tab:XRT}). \textit{Bottom}: \swift/BAT light curve of \target in the 15--50\,keV band. The blue arrows mark epochs of \nustar observations (Table~\ref{tab:nustar}). The dashed vertical line in all three panels marks the optical first detection epoch.\label{fig:xlc}}
\end{figure*}

\begin{figure*}[htbp!]
	\centering
	\includegraphics[width=\textwidth]{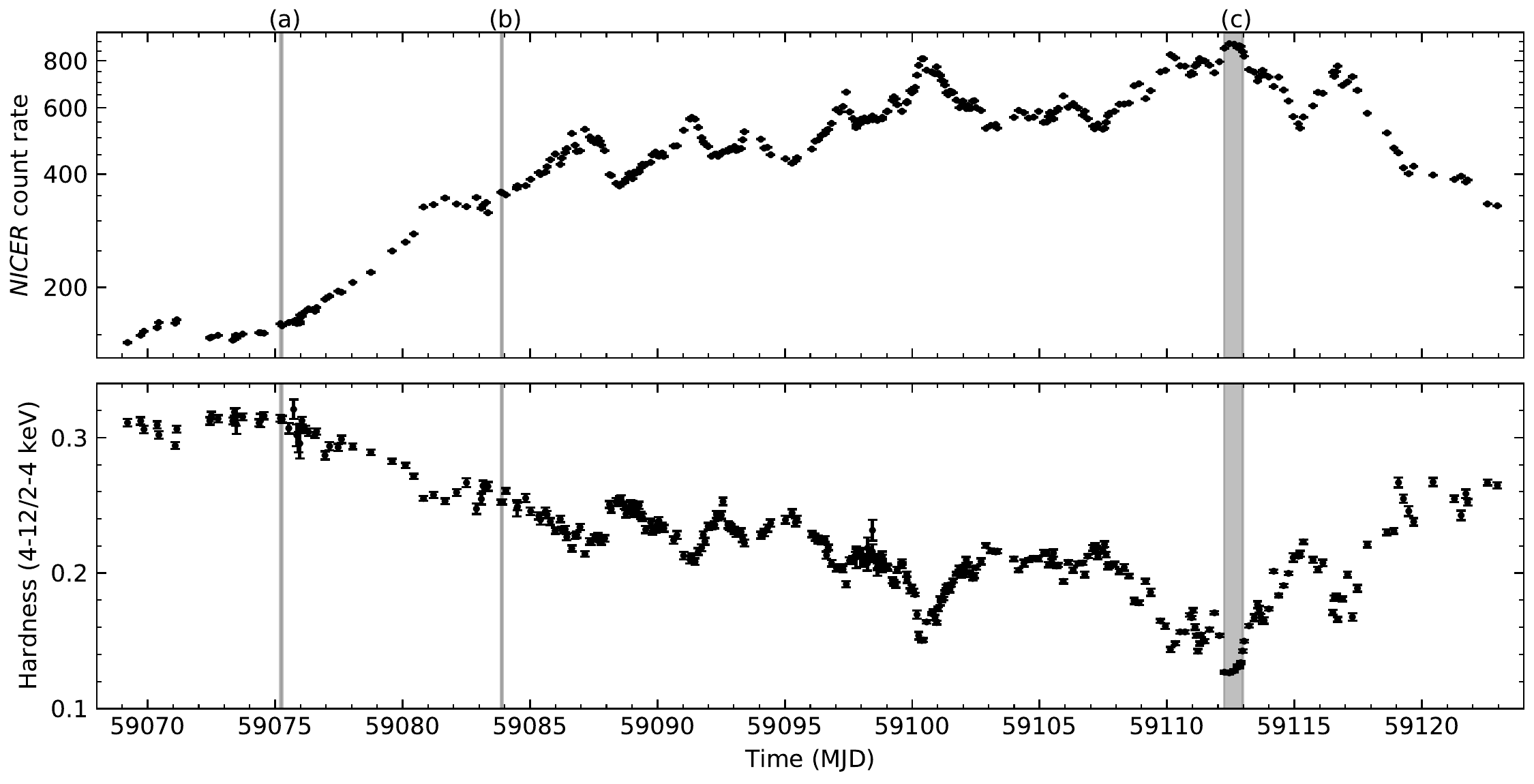}
	\caption{\textit{Upper}: \nicer (0.4--12\,keV) light curve of \target.
	\textit{Bottom}: \nicer hardness (ratio of the 4--12\,keV and 2--4\,keV count rates) of \target. %is shown in blue (upward triangles) and soft color (ratio of the 1--2\,keV and 0.4--1\,keV count rates) is shown in red (downward triangles). 
	The three vertical grey regions mark epochs where detailed timing analysis are performed (see Section~\ref{subsubsec:nicer_aperiodic} and Table~\ref{tab:pds_param}.)
	\label{fig:nicer_lc}}
\end{figure*}

\subsection{\maxi}
\maxi was installed on the Japanese Experiment Module Exposed Facility on the ISS in 2009 July. Since 2009 August, the \maxi Gas Slit Cameras (GSCs; \citealt{Mihara2011, Sugizaki2011}) have been observing the source region of \target in the 2--20\,keV band every 92\,min synchronized with the ISS orbital period. Owing to the ISS orbit precession of about 72\,days, the source region, due to the interference of some structure of the detectors, has been regularly unobservable for about 12\,days. Furthermore, in recent years, the source has only been observed with the degraded cameras for $\sim28$\,days in each precession period. We did not use these data. As a result, there are two data gaps every 72\,days.

The 1-day average \maxi light curves were generated by the point-spread-function fit method \citep{Morii2016} to obtain the most reliable curves in the 2--4\,keV and 4--10\,keV bands. Furthermore, we excluded data with 1$\sigma$ uncertainties 2.5 times larger than the average uncertainties in the 2--4 and 4--10\,keV bands, respectively. We rebinned the data into 4-day bins to improve the statistics. 

\subsection{\swift}
\subsubsection{XRT}

AT2019wey was observed by the X-Ray Telescope (XRT; \citealt{Burrows2005}) on board \swift in 2020 April (4 epochs), August (5 epochs), and September (4 epochs). We generated the X-ray light curve for AT2019wey using an automated online tool\footnote{\url{https://www.swift.ac.uk/user_objects}} \citep{Evans2009}. The first 9 epochs were obtained in Photon Counting (PC) mode, and thus suffer from ``pile-up'' at the high observed count rates. Standard corrections \citep{Evans2007} were applied to the observations taken in 2020 April. The observations from 2020 August were sufficiently piled up that no reliable count rates could be obtained. Beginning in 2020 September, XRT observations were obtained in Window Timing (WT) mode, where larger count rates can still be reliably measured. The observation log and the count rate measurements are shown in Table~\ref{tab:XRT}.

To generate XRT spectra for the 4 epochs obtained in 2020 April, we processed the data using \texttt{xrtproducts}. We extracted source and background photons from circular regions with radii of 50$^{\prime\prime}$ and 100$^{\prime\prime}$, respectively.

\subsubsection{BAT}\label{subsubsec:bat}
Since 2004 November, the source region of \target has been observed by the Burst Alert Telescope (BAT; \citealt{Krimm2013}) on board \swift. The 15--50\,keV BAT light curve is provided by the ``scaled map'' data product\footnote{Available at \url{https://swift.gsfc.nasa.gov/results/transients/weak/AT2019wey/}.}. To generate a light curve with better statistics, we rebinned the light curve into 4-day bins, and excluded data with 1$\sigma$ uncertainties 3 times larger than the median uncertainties.

\begin{figure*}[htbp!]
	\centering
	\includegraphics[width=0.8\textwidth]{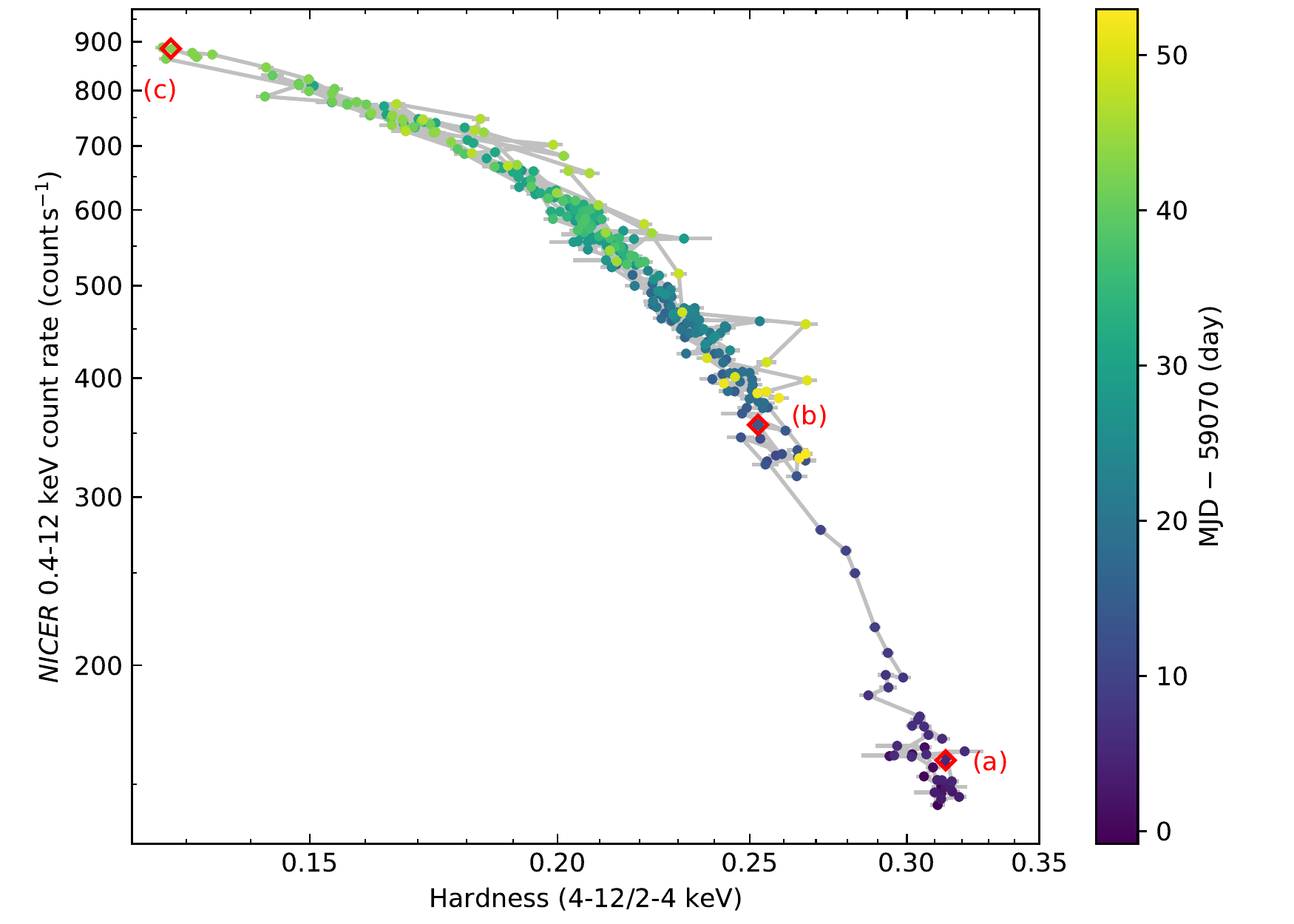}
	\caption{The \nicer HID, defined as the 0.4--12\,keV count rate versus the hardness (ratio of the 4--12\,keV and 2--4\,keV count rates). The data points are color coded by time. The three red diamonds mark epochs where detailed timing analysis are performed (see Section~\ref{subsubsec:nicer_aperiodic} and Table~\ref{tab:pds_param}).
	\label{fig:nicer_hid}}
\end{figure*}

\begin{figure}[htbp!]
    \centering
    \includegraphics[width=\columnwidth]{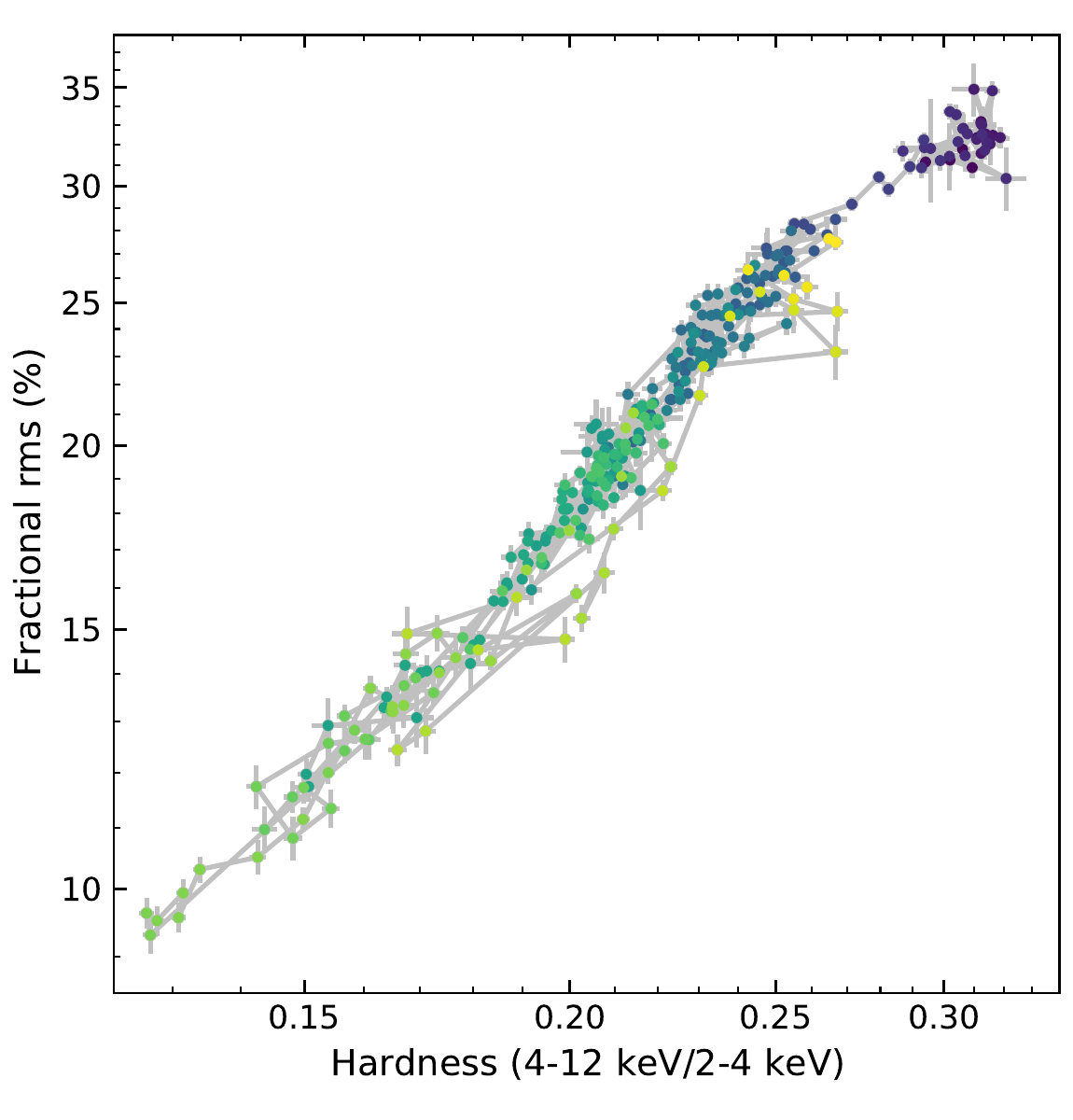}
    \caption{The \nicer HRD, defined as the hardness versus the fractional rms integrated in the 0.1--64\,Hz frequency range. The data is color-coded by time following the scale shown in Figure~\ref{fig:nicer_hid}.}
    \label{fig:nicer_hrd}
\end{figure}

\begin{figure}[htbp!]
    \centering
    \includegraphics[width=\columnwidth]{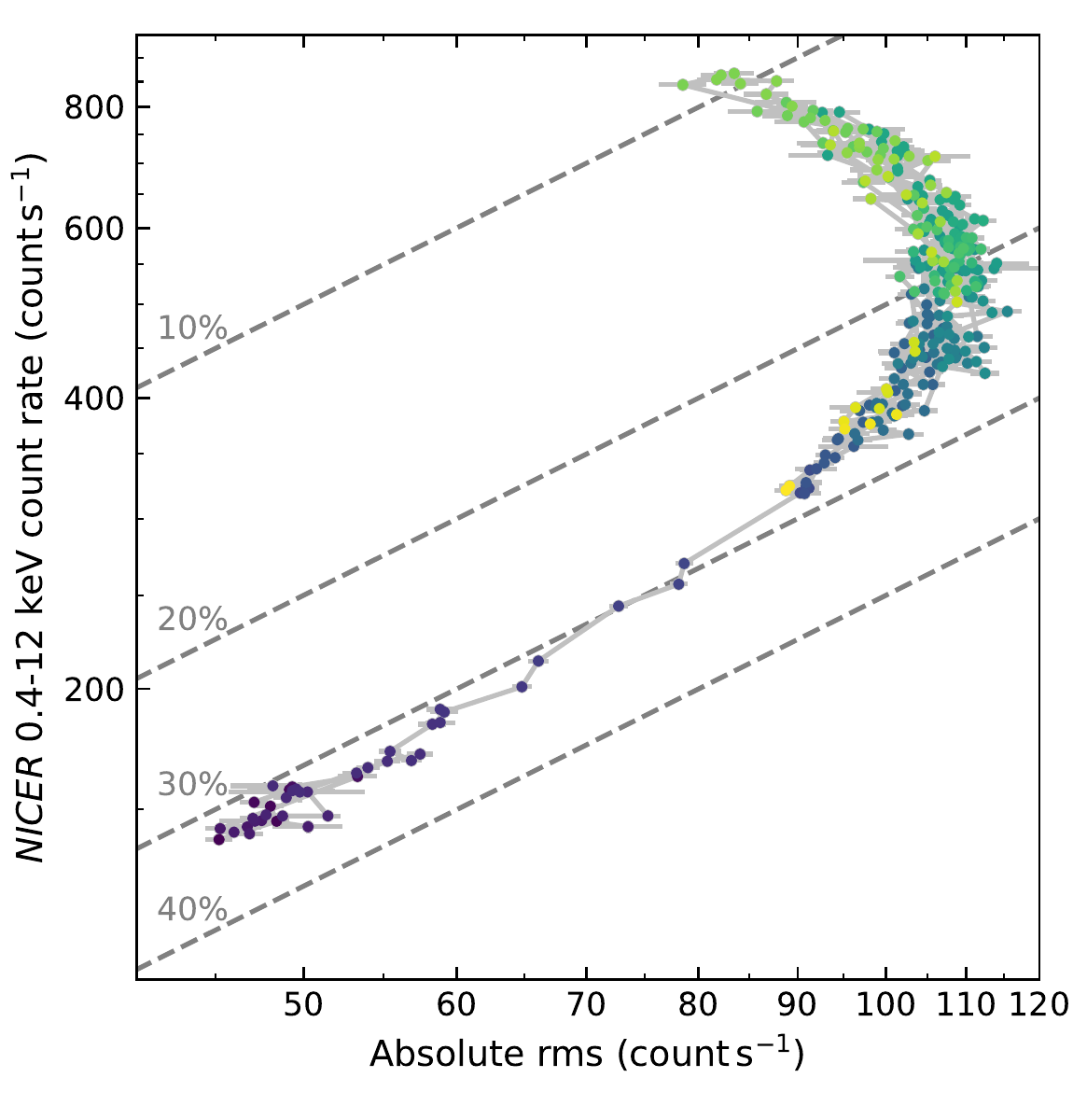}
    \caption{The \nicer RID, defined as the absolute rms versus the 0.4--12\,keV count rate. The gray dotted lines represent the 10, 20, 30 and 40 percent fractional rms levels. The data is color-coded by time following the scale shown in Figure~\ref{fig:nicer_hid}.}
    \label{fig:nicer_rid}
\end{figure}

\section{Analysis of Light Curves} \label{sec:result}
The \maxi, \nicer, and \swift/BAT light curves of \target are shown in Figure~\ref{fig:xlc}. The dashed vertical lines in the three panels mark the epoch of first optical detection on 2019 December 2 (Paper II). From 2019 January 1 to December 2, the \maxi/GSC and \swift/BAT data show no significant count excess. We refer to \citet{Hori2018} for \maxi/GSC detection upper limits during the period from 2009 August 13 to 2016 July 31. From the 2019 December 2 to the \srg discovery epoch (2020 March 18), \maxi/GSC detected a significant 2--10\,keV flux excess of $1.7\pm0.4$\,mCrab (see \citealt{Negoro2020}), and BAT detected a significant 15--50\,keV flux excess of $3.7\pm0.7$\,mCrab.

As can be seen from the \maxi and BAT light curves, the source started to significantly brighten from the beginning of 2020 June to the middle of 2020 August. Since then, the source has stayed at a relatively high level of flux. From 2020 September 2 to 2020 November 30, the median \maxi 2--10\,keV flux is 17.7\,mCrab and the median BAT 15--50\,keV flux is 28.1\,mCrab. The \nicer light curve is presented in the upper panel of Figure~\ref{fig:nicer_lc}. It clearly shows that after the X-ray brightening, \target underwent a few week-long mini-outbursts in the 0.4--12\,keV band.

\subsection{Hardness Evolution} \label{subsec:colors}
We define the X-ray hardness (or X-ray color) using the ratio of count rates in the \nicer 4--12\,keV and 2--4\,keV bands. The evolution of hardness is shown in the bottom panel of Figure~\ref{fig:nicer_lc}. Figure~\ref{fig:nicer_hid} presents the \nicer HID of \target. At the beginning, the source was faint and hard. As it got brighter, the X-ray color became softer. The count rate (hardness) reached the maximum (minimum) value at 59112\,MJD, after which the count rate decreased and the X-ray color hardened. The evolution of \target roughly follows a single line on the HID, i.e., each hardness value corresponds to a single value of count rate. This is very different from the hysteresis pattern generally observed in LMXBs.

\subsection{Timing Properties} \label{subsec:timing}

The typical event timestamps for \nicer/XTI and \nustar are accurate to $\sim100$\,ns \citep{Prigozhin2016} and $\sim100$\,$\mu$s \citep{Bachetti2021}, respectively. The high timing precision makes the two instruments ideal to study fast X-ray variability. We searched for coherent pulsation signals in the \nicer and \nustar data and found no viable pulse search candidates to 3$\sigma$ level despite pulsation searches extending to 100\,ns (see Appendices \ref{subsec:nicer_pulsations} and \ref{subsec:nustar_pulsations} for details). Here, we present aperiodic analysis of \nicer (Section~\ref{subsubsec:nicer_aperiodic}) and \nustar (Section~\ref{subsubsec:nustar_aperiodic}) observations. 

\subsubsection{\nicer Aperiodic Analysis} \label{subsubsec:nicer_aperiodic}

We produced an average PDS in the 0.5--12\,keV energy band for each GTI. We used 16-s long intervals and $\approx0.12$\,ms time resolution. The average PDS was rms-normalized \citep{Belloni1990} and the contribution due to the photon counting noise was subtracted. We calculated the integrated fractional rms in the 0.1--64\,Hz frequency range. We also calculated the absolute rms by multiplying the fractional rms by the net count rate \citep{Munoz-Darias2011}.

In Figure \ref{fig:nicer_hrd} we show the hardness--fractional rms diagram (HRD), which is usually used to study the outburst evolution of transient BH LMXBs \citep{Belloni2005}. The integrated fractional rms decreased from $\sim$30\% to $\sim$10\% as the X-ray color softened, and increased back to $\sim$25\% as the color hardened again. 

In Figure \ref{fig:nicer_rid} we show the absolute rms--intensity diagram (RID). At the beginning of the X-ray brightening, we found that the absolute rms increased with the count rate. This linear trend has been observed in many BH binaries, and is commonly known as the ``hard line'' (HL; \citealt{Munoz-Darias2011}). Starting from $\sim$59082\,MJD, the source left the HL and moved upwards. During the light curve bumps observed between $\sim$59085\,MJD and $\sim$59123\,MJD, the source moved to the left as the count rate increased, and then went back as the count rate decreased.

\begin{figure*}[htbp!]
	\centering
	\includegraphics[width=\textwidth]{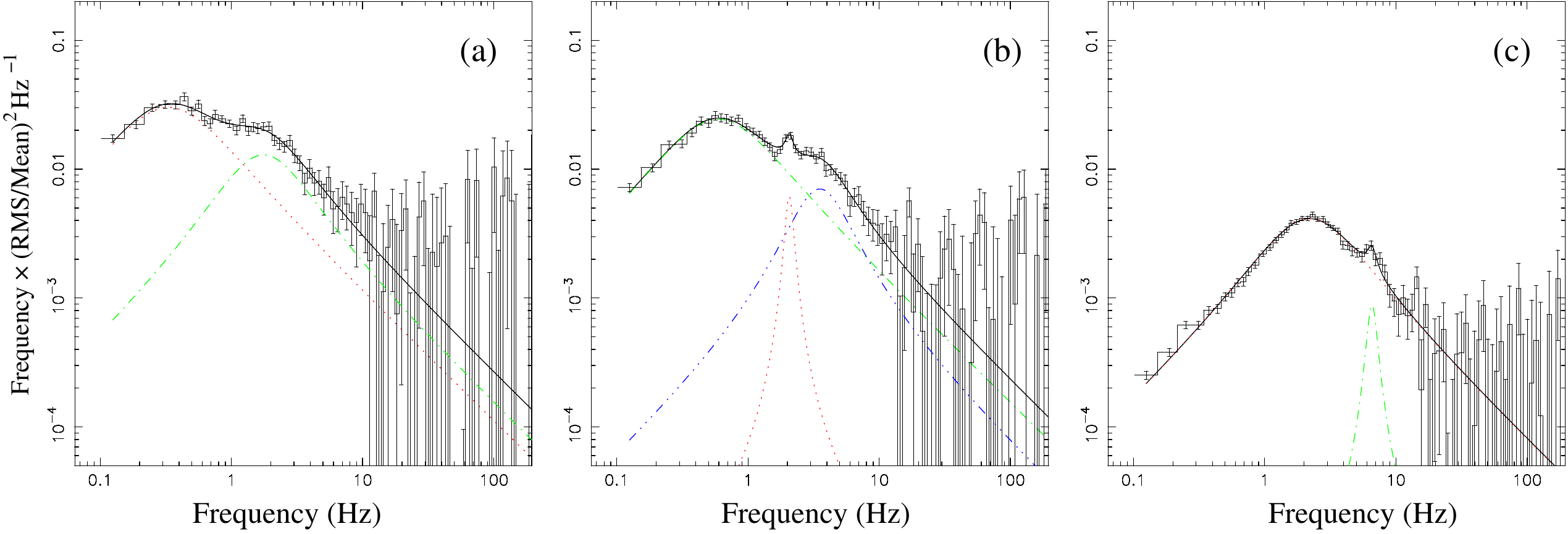}
	\caption{Representative \nicer power spectra. The power spectra were calculated in the 0.5--12\,keV energy band. The main properties of the power spectra are listed in Table \ref{tab:pds_param}. The PDS is fitted with a combination of two or three Lorentzian functions, as shown by the colored components in each panel.
	\label{fig:nicer_powerspec}}
\end{figure*}

During the period we analyzed, the PDS can be well fitted with two or three Lorentzian functions following the prescription laid out by \citet{Belloni2002}.
In Figure \ref{fig:nicer_powerspec} we show three representative PDS averaged from different phases of the outburst (marked as grey regions in Figure~\ref{fig:nicer_lc} and red diamonds in Figure~\ref{fig:nicer_hid}). 

\begin{deluxetable}{ccccc}[htbp!]
\tablecaption{\nicer power spectral components\label{tab:pds_param}}
\tablehead{
\colhead{TIME (MJD)}  
& \colhead{} 
& \colhead{$\nu_{\mathrm{max}}$ (Hz)} 
& \colhead{$Q$} 
& \colhead{rms (\%)}}
\startdata
\multirow{2}{*}{59075.20--59075.29}     & $L_1$      & $0.33 \pm 0.02$  & $0.25 \pm 0.08$  & $27.57 \pm 1.12$  \\
                                        & $L_2$      & $1.76 \pm 0.14$  & $0.44 \pm 0.12$  & $16.33 \pm 1.62$  \\
\hline
\multirow{3}{*}{59083.85--59083.94}     & $L_1$      & $0.59 \pm 0.02$  & $0.26 \pm 0.04$  & $24.62 \pm 0.34$  \\
                                        & $L_2$      & $2.06 \pm 0.03$  & 6 (fixed)        & $3.95  \pm 0.41$  \\
                                        & $L_3$      & $3.53 \pm 0.16$  & $0.81 \pm 0.14$  & $10.17 \pm 0.69$  \\
\hline
\multirow{2}{*}{59112.24--59112.98}     & $L_1$      & $2.21 \pm 0.03$  & $0.34 \pm 0.02$  & $9.72 \pm 0.07$  \\
                                        & $L_2$      & $6.58 \pm 0.21$  & $4.99 \pm 1.97$  & $1.64 \pm 0.26$  \\
\enddata
\end{deluxetable}

The main properties of the PDS are listed in Table \ref{tab:pds_param}.
At the beginning of the outburst, the PDS were dominated by strong band-limited noise without showing any significant QPOs. The average PDS can be fitted with two broad Lorentzians (Figure \ref{fig:nicer_powerspec} (a)).
Starting from $\sim$59083\,MJD, a weak QPO was sometimes observed in the PDS. The characteristic frequency of the QPO increased from $\sim$2\,Hz to $\sim$6.5\,Hz as the spectra softened. 
Figure~\ref{fig:nicer_powerspec} (b)--(c) show the PDS of the QPO with the lowest and highest frequency, respectively.
Based on the properties of the QPO and noise, this QPO is similar to the type-C QPO \citep[e.g.][]{Casella2005, Motta2011, Ingram2019, Zhang2020} commonly observed in BH and NS binaries \citep[see, e.g.,][]{Klein-Wolt2008}.

\subsubsection{\nustar Aperiodic Analysis} \label{subsubsec:nustar_aperiodic}
\begin{figure*}[htbp!]
\begin{center}
\includegraphics[width=0.45\textwidth]{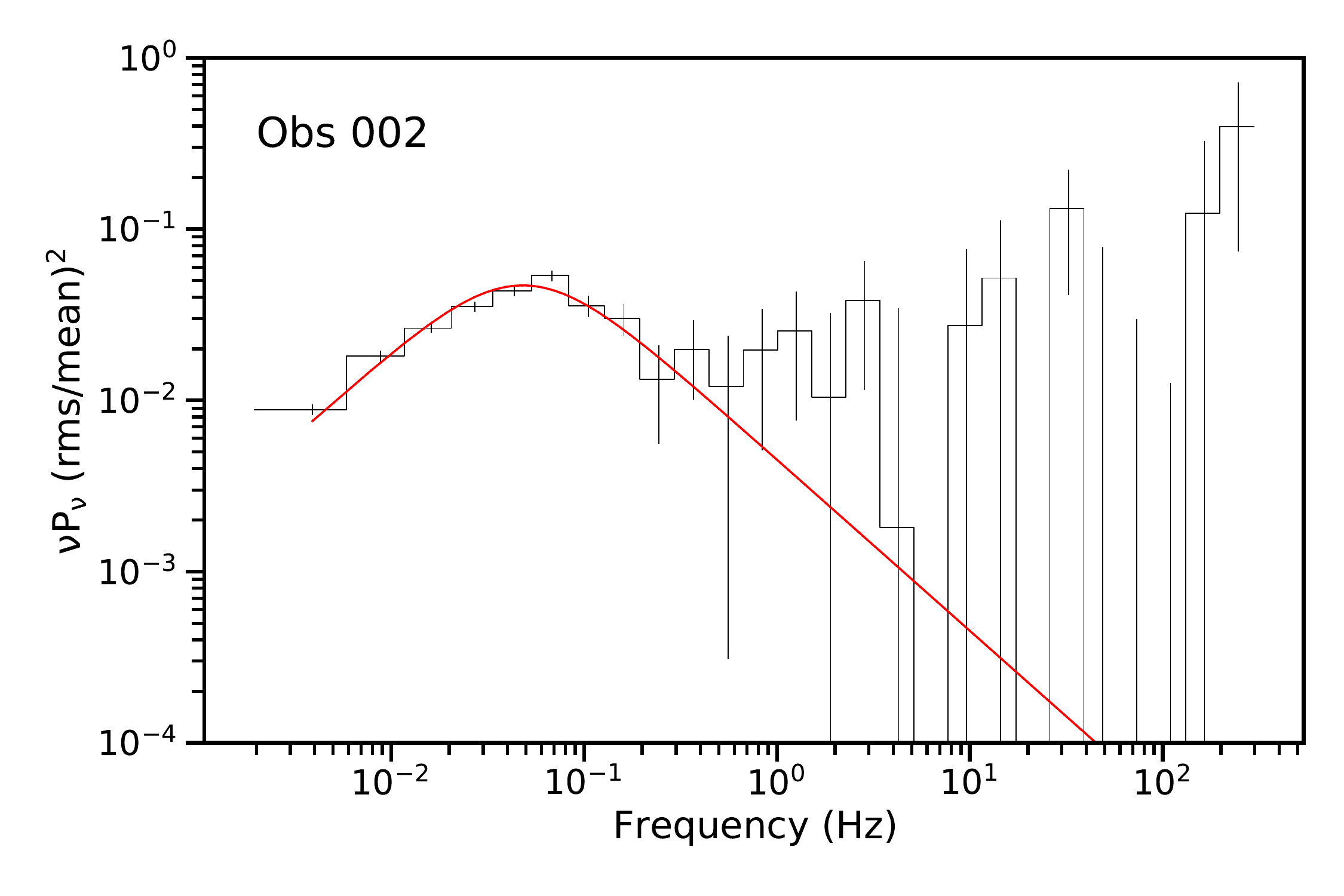}
\includegraphics[width=0.45\textwidth]{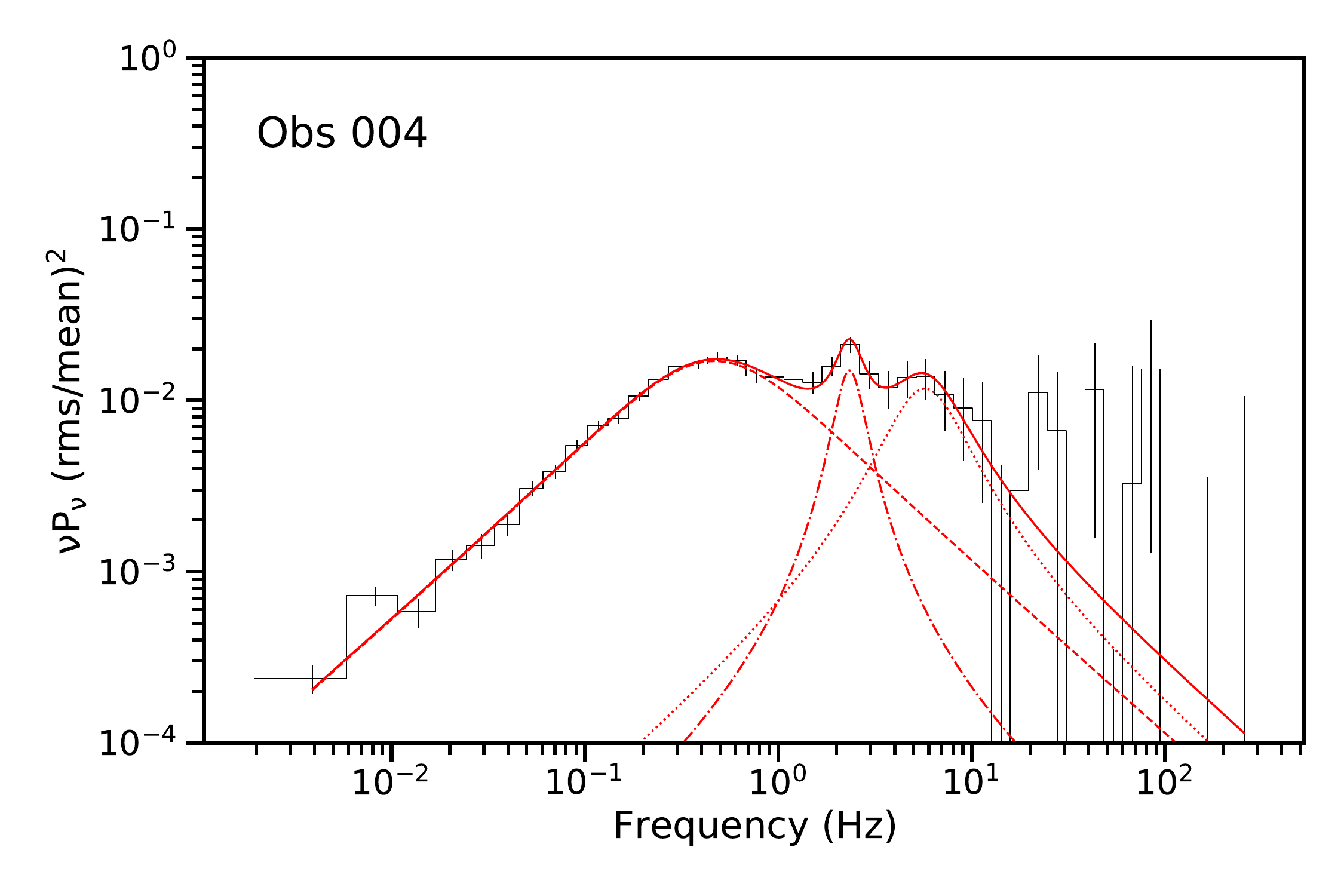}
\caption{The averaged rms-normalized cospectra for each \nustar observation are shown in black in units of $\mathrm{Power \times Frequency}$. Their best-fit models are plotted in solid red and the individual Lorentzian components, $L_b$, $L_\mathrm{LF}$, and $L_h$, are shown in dashed, dashed-dotted, and dotted red, respectively. Cospectra have been rebinned for legibility.
\label{fig:cospectra_nuPnu}}
\end{center}
\end{figure*}

Rather than summing the FPMA and FPMB light curves and producing PDS for each observation, we chose to analyze the Cross Power Density Spectrum (CPDS; \citealt{Bachetti2015}). The CPDS taken between FPMA and FPMB is given by
\begin{equation}
    C(\nu) = \mathcal{F}^{*}_{\mathrm{A}}(\nu) \mathcal{F}_{\mathrm{B}}(\nu)
    \label{eq:CPDS}
\end{equation}
where $\mathcal{F}^{*}_{\mathrm{A}}(\nu)$ is the complex conjugate of the Fourier transform of the light curve observed by FPMA and $\mathcal{F}_{\mathrm{B}}(\nu)$ is the Fourier transform corresponding to FPMB. 
%The CPDS between two light curves gives the relationship between their variability properties. 
The real part of the CPDS, called the cospectrum, represents only the power of the signals which are in phase between the two light curves, and its imaginary part gives the power of those signals which are 
%out of phase. 
in quadrature. The CPDS can therefore be used to calculate time lags and correlations between two light curves. 
% It is often used to compare the variability between different energy bands as measured by a single instrument. 
%In addition, these properties make the CPDS particularly useful for analyzing \nustar data. Whereas the PDS requires careful noise subtraction to account for dead time and Poisson noise, the cospectrum calculated between FPMA and FPMB brings the average white noise level down to 0 since these effects are not correlated between the two instruments. 
%For more information about the CPDS, the cospectrum, and its applications for \nustar timing analysis, see \citet{Bachetti2015}.

\begin{figure*}[htbp!]
\begin{center}
\includegraphics[width=0.45\textwidth]{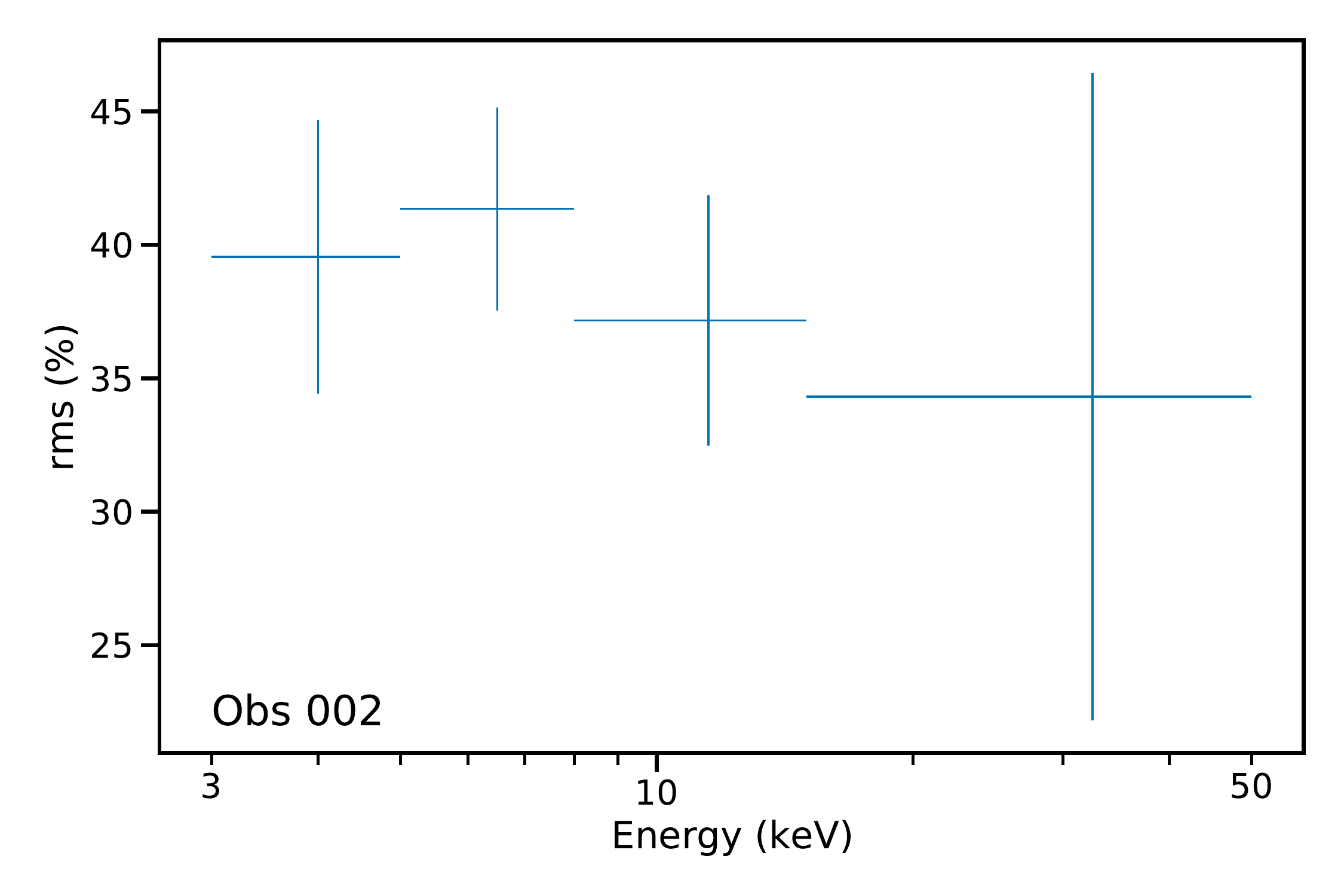}
\includegraphics[width=0.45\textwidth]{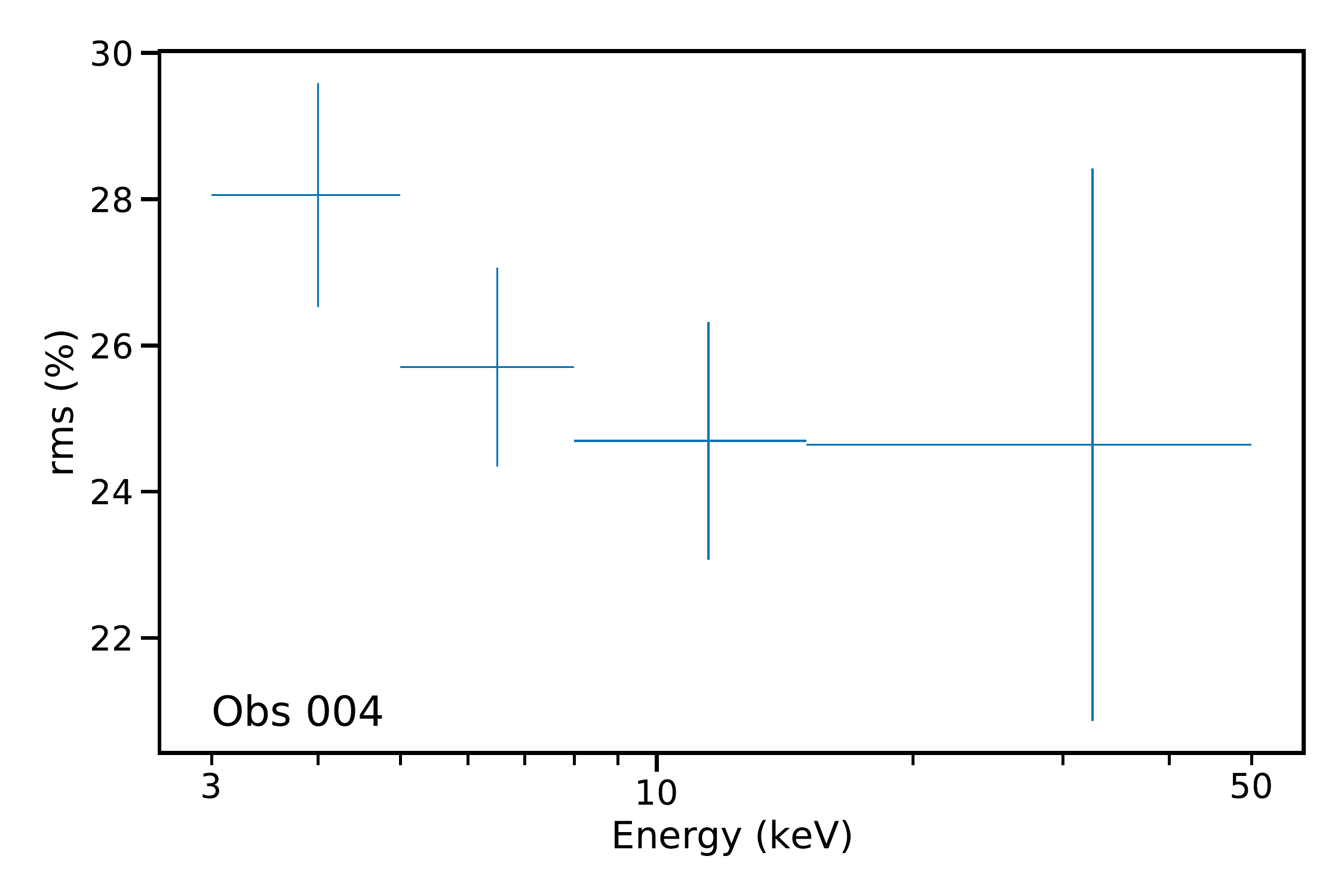}
\caption{The observed variability as measured by the fractional rms is shown as a function of photon energy for \nustar observations. While observation 002 is consistent with a flat rms-energy relation, observation 004 shows evidence of decreasing variability with increasing energy.
\label{fig:rms_E}}
\end{center}
\end{figure*}

In order to produce a cospectrum for each observation, we split the light curves observed by each FPM into intervals of 256\,s each, resulting in 150 intervals for observation 002 and 173 intervals for observation 004. For each of these intervals, we produced a cospectrum, and then averaged these cospectra together. The frequencies sampled are limited to the range $ 4\,\mathrm{mHz} < \nu < 256\,\mathrm{Hz}$. The low end of this range is determined by the interval length, and the high end is determined by the sampling rate of the light curves. The resulting averaged, rms-normalized cospectra for observations 002 and 004 are shown in black in Figure \ref{fig:cospectra_nuPnu}, where they have been rebinned for clarity. All errors quoted are 1-$\sigma$.

%\subsubsection{Modeling the Power Spectra}\label{subsubsec:modeling}
Similar to our analysis in Section~\ref{subsubsec:nicer_aperiodic}, we fit the cospectra with a model consisting of a sum of Lorentzian functions following \citet{Belloni2002}. We used an automated modeling algorithm that fits a cospectrum to composite Lorentzian models with progressively more components, halting when the addition of a component no longer results in the reduction of the $\chi^{2}$ fit statistic. We chose the model with the minimum number of components which still resulted in a significant improvement to the fit ($|\Delta\chi^{2}| > 10$), and discarded more complex models with only marginally better fit statistics. For observation 002, this resulted in a single-component model containing only one broad Lorentzian with unconstrained $\nu_{0}$ and $Q$. For observation 004, we obtained a model with two broad components centered at considerably higher frequencies than that of the component obtained for observation 002. Following the notation of \citet{Klein-Wolt2008}, we dub the lowest frequency broad components $L_{b}$, and the higher frequency broad component observed in observation 004 $L_{h}$. 

\begin{deluxetable}{ccccc}[htbp!]
\tablecaption{\nustar power spectral components.\label{tab:cospec_param}}
\tablewidth{0pt}
\tablehead{
\colhead{OBSID}  & \colhead{Component} & \colhead{$\nu_{\mathrm{max}}$ (Hz)} & \colhead{$Q$} & \colhead{rms (\%)}}
\startdata
90601315002                     & $L_{b}$   & $0.05$\tablenotemark{$\dagger$} & $3\times 10^{-4}$\tablenotemark{$\dagger$} & $54 \pm 5$ \\
\hline
\multirow{3}{*}{90601315004}    & $L_{b}$   & $0.5 \pm 0.1$     & $0.15 \pm 0.03$       & $28 \pm 1$  \\
                                & $L_{h}$   & $5.7 \pm 1.4$     & $0.9 \pm 0.6$         & $14 \pm 3$  \\
                                & $L_{\mathrm{LF}}$       & $2.3 \pm 0.1$     & $2.6 \pm 1.0$         & $9.5 \pm 1.9 $  \\
\enddata
\tablenotetext{\dagger}{The characteristic frequency and quality factor are not constrained for observation 002, therefore errors are not shown for these quantities.}
\end{deluxetable}

Following the detection of the two broad components in observation 004 using our fitting algorithm, visual inspection suggested the presence of an additional component at $\sim2\,\mathrm{Hz}$. We therefore added a third QPO-like component to the model and saw a small but significant improvement to the fit of $\Delta\chi^{2} = -30$. We label this narrower QPO-like component $L_{\mathrm{LF}}$. Defining the QPO significance as the ratio of the integrated power of the component to its error, $A/\sigma_{A}$, the significance of $L_{\rm LF}$ was calculated to be 2.5$\sigma$. Note that this component lines up with the QPO seen in the \nicer PDS (Figure~\ref{fig:nicer_powerspec} (b)), and it is therefore still significant. All of the components observed in each observation as well as their fitted parameters are listed in Table \ref{tab:cospec_param}. The components and the resulting composite models are shown in red in Figure~\ref{fig:cospectra_nuPnu}.

Finally, in order to better understand the physical origins of the source variability, we computed the variability as a function of photon energy for each observation. We produced cospectra in four energy ranges, and determined the fractional rms by integrating the cospectra. Due to the limited frequency range for which significant power was detected, we did not integrate over the entire available frequency range. Rather, for observation 002, we integrated the power between 4\,mHz and 1\,Hz, while for observation 004, we integrated the power between 4\,mHz and 10\,Hz. The resulting rms-energy relations are shown in Figure \ref{fig:rms_E}. Observation 002 is consistent with a flat rms-energy relation, whereas observation 004 may exhibit decreasing variability with increasing photon energy. 

\begin{figure*}[htbp!]
\centering
\includegraphics[width=0.8\textwidth]{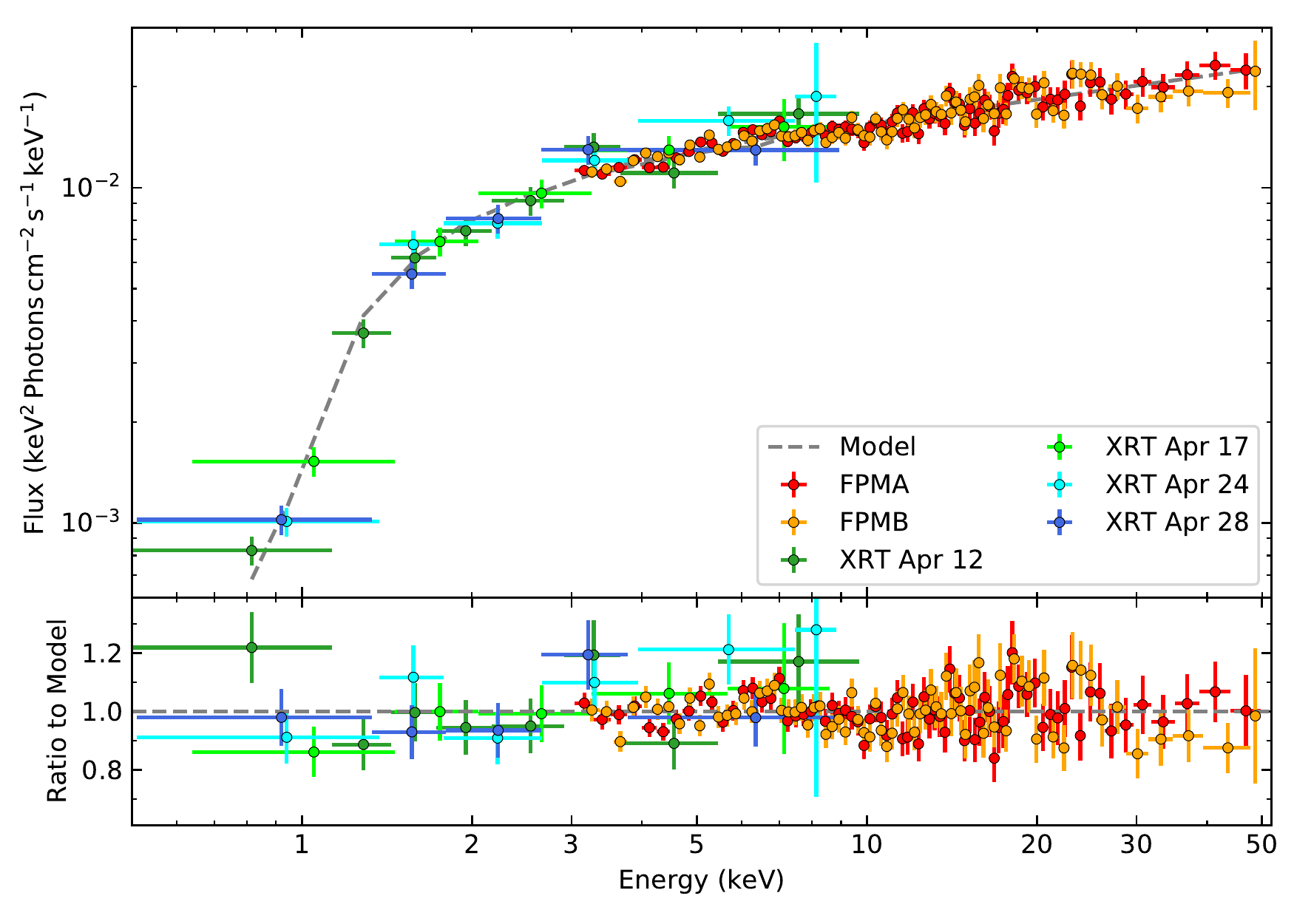}
\caption{\textit{Upper}: Unfolded spectrum of the data from the two \nustar detector modules and the four \swift/XRT observations. The best-fit model for \nustar-FPMA is shown in the dashed grey line. To account for the cross-normalization terms (see Table~\ref{tab:april}), the FPMB and XRT data are divided by 1.037 and 0.796, respectively. \textit{Bottom}: The ratio of data to the best-fit model is shown for all three data sets. The data have been rebinned for visual clarity.
\label{fig:april_spec}
}
\end{figure*}

\section{Spectral Analysis} \label{sec:spectral}

In this section, we examine the spectral evolution of \target. For this analysis we use  \texttt{xspec} version 12.11.0 \citep{Arnaud1996}. Uncertainties of model parameters are represented by the 90\% confidence intervals, which are estimated by the \texttt{error} command in \texttt{xspec}. 

Below we present joint analysis (i.e., analysis of contemporaneous datasets obtained from several missions) and also specific data sets.

In Section~\ref{subsec:spectral_joint1}--\ref{subsec:spectral_joint3}, we perform joint spectral analyses of three sets of observations obtained in 2020 April, August, and September. 
Section ~\ref{subsec:spectral_joint1} presents the April 2020 epoch, where the \nustar 002 spectra for FPMA and FPMB were fitted with data from four \swift/XRT observations obtained in April 2020 (Table~\ref{tab:XRT}).
Section ~\ref{subsec:spectral_joint2} presents the August 2020 epoch, where the \nustar 004 spectra for FPMA and FPMB were fitted with two \nicer observations bracketing the \nustar observation (Table~\ref{tab:nicer}, OBSID 3201710112 and 3201710113).
Section ~\ref{subsec:spectral_joint3} presents the September 2020 epoch, where the \chandra spectra were fitted with two \nicer observations bracketing the \chandra observation (Table~\ref{tab:nicer}, OBSID 3201710147 and 3201710148).
In Section~\ref{subsec:spectral_nicer}, we analyze the \nicer spectra for each OBSID between 3201710105 and 3201710157. 

\begin{deluxetable}{lr}[htbp!]
\tablecaption{Best-fit model parameters of the 2020 April joint observations.  \label{tab:april} }
\tablehead{
\colhead{Parameter \hspace{1.8cm} }  
& \colhead{ \hspace{1.8cm} 90\% Interval} }
\startdata
\multicolumn{2}{c}{\texttt{constant}} \\ \hline
$\mathcal{C_{\rm FPMA}}$ & 1 (frozen)\\
$\mathcal{C_{\rm FPMB}}$ & $ 1.037\pm 0.014$\\% checked
$\mathcal{C_{\rm XRT}}$ & $0.796_{-0.038}^{+0.039}$\\% checked
\multicolumn{2}{c}{\texttt{tbabs}} \\ \hline
$N_{\rm H}$ (10$^{22}$\,\cmcm) & $0.609_{-0.045}^{+0.049}$ \\
\multicolumn{2}{c}{\texttt{powerlaw}} \\ \hline
$\Gamma$ & $1.765 \pm 0.013$ \\
\texttt{norm}\tablenotemark{$\dagger$} &  $9.06_{-0.26}^{+0.27}$  \\ \hline
\texttt{C-stat / d.o.f.} & 1292.70/1541 \\ \hline
\enddata
\tablenotetext{\dagger}{Normalization at 1\,keV in units of ${\rm 10^{-3}\,ph\,keV^{-1}\,cm^{-2}}$.}
\end{deluxetable}

\subsection{Joint Analysis, 2020 April}  \label{subsec:spectral_joint1}
The upper panel of Figure~\ref{fig:april_spec} shows the \nustar 002 and \swift/XRT spectra in April 2020, which appears relatively featureless. We therefore modeled the data with an absorbed power-law (\texttt{tbabs*powerlaw}, in \texttt{xspec}, \citealt{Wilms2000}). We also included a leading cross-calibration term (\texttt{constant}; \citealt{Madsen2017}) between the two \nustar telescopes (with $\mathcal{C_{\rm FPMA}}$ defined to be 1) and a single 
term used for all four \swift/XRT observations. 
The four XRT spectra were grouped with \texttt{grppha} to have at least one count per bin. The \nustar FPMA and FPMB spectra were grouped with \texttt{ftgrouppha} using the optimal binning scheme developed by \citet{Kaastra2016}. All data were fitted using $C$-statistics via \texttt{cstat} \citep{Cash1979}. For \nustar we fitted the data over the 3--50\,keV range as the source spectrum becomes comparable to the background at higher energies, while for \swift we fitted from 0.5 to 10\,keV.

The best-fit model is shown in Figure~\ref{fig:april_spec}. The model parameters are given in Table \ref{tab:april}. We note that the $\mathcal{C_{\rm XRT}}$ is lower than we would typically expect. A probable explanation is that the pile-up resulted in an observed flux lower than expected, as the source count rate is relatively high for the XRT PC mode (see Table~\ref{tab:XRT}).

The unabsorbed flux in the 0.3--100\,keV band for FPMA is $1.4\times 10^{-10}\,{\rm erg\,cm^{-2}\,s^{-1}}$. Paper II constrains the distance of \target to be $1 \lesssim D \lesssim 10$\,kpc. At distances of [1, 3, 5, 10]\,kpc, this corresponds to a luminosity of [0.16, 1.5, 4.1, 16.3]$\times 10^{35}\,{\rm erg\,s^{-1}}$. The Eddington luminosity is $L_{\rm Edd} = 1.46\times 10^{38} (M/M_\odot)\,{\rm erg\,s^{-1}}$ (assuming solar hydrogen mass fraction $X=0.71$). Therefore, the X-ray luminosity in 2020 April is $10^{-5} \lesssim L_{\rm X}/L_{\rm Edd} \lesssim  10^{-3}$ for a $\approx$10\,$M_\odot$ compact object.

\begin{deluxetable}{lr}[htbp!]
\tablecaption{Best-fit model parameters of the 2020 August 16 joint observations. \label{tab:august} }
\tablehead{
\colhead{Parameter \hspace{2cm}}  
& \colhead{\hspace{2cm} 90\% Interval} }
\startdata
\multicolumn{2}{c}{\texttt{constant}} \\ \hline
$\mathcal{C_{\rm FPMA}}$ & 1 (frozen)\\ % explained
$\mathcal{C_{\rm FPMB}}$ & $ 1.051\pm 0.003$\\% explained
$\mathcal{C_{\nicer}}$ & $1.035 \pm 0.002$\\% explained
\multicolumn{2}{c}{\texttt{tbfeo}} \\ \hline
$N_{\rm H}$ (10$^{22}$ \cmcm) & $0.513\pm 0.003$ \\ %checked
O & $<0.020$ \\ %checked
Fe & $<0.0528$ \\%checked
$z$ & 0 (frozen) \\%checked
\multicolumn{2}{c}{\texttt{simplcutx}} \\ \hline
$\Gamma$ & $1.786 \pm 0.001$  \\ % explained
$f_{\rm sc}$ & $0.746 \pm 0.005$\\ % explained
$R_{\rm F}$ & 1 (frozen)\\ % explained
$kT_e$ (keV) & 1000 (frozen)\\ % explained
\multicolumn{2}{c}{\texttt{diskbb}} \\ \hline
$T_{\rm disk}$ (keV) & $0.3542 \pm 0.0001$ \\% checked
%norm$_{\rm disk}^{a}$ &  $1647_{-3}^{+2}$ \\ % checked
R$_{\rm in}^{\ast}$\tablenotemark{$\dagger$} &  $40.58\pm0.03$ \\ % checked
\multicolumn{2}{c}{\texttt{relxillCp}} \\ \hline
$q$ & 3 (frozen)\\ % explained
%$R_{\rm break}$ ($R_{\rm isco}$) & 1 (frozen)\\ % explained
$a$ & 0 (frozen)\\ % explained
$i$ (deg) & $27.0_{-1.2}^{+0.8}$\\ % explained
$R_{\rm in}$ ($R_{\rm isco}$) &  $<1.05$\\% explained
$R_{\rm out}$ ($R_{\rm g}$) &  400 (frozen)\\% explained
log$\xi$ & $3.0121^{+0.0016}_{-0.0020}$ \\ % explained
$A_{\rm Fe}$ & $2.86_{-0.09}^{+0.10}$ \\ % explained
$kT_e$ (keV) & 1000 (frozen)\\ % explained
$R_{\rm F}$ & 1 (frozen)\\ % explained
Norm$_{\rm rel}$ ($10^{-4}$) & $2.96 \pm 0.03$\\% explained
\multicolumn{2}{c}{\texttt{edge}} \\ \hline
$E_{\rm c}$ (keV) & $1.369^{+0.017}_{-0.016}$ \\% checked
$D$ & $ 0.071^{+0.005}_{-0.005}$ \\  % checked
\hline
\texttt{C-stat / d.o.f.} & 2006.52 (1769) \\
\enddata
\tablenotetext{\dagger}{Normalization $(R_{\rm in}/D_{10})\sqrt{{\rm cos}i}$, where $R_{\rm in}$ is the inner disk radius in the unit of km, and $D_{10}$ is distance to the source in units of 10\,kpc.}
\end{deluxetable}

\begin{figure*}[htbp!]
\centering
\includegraphics[width=0.8\textwidth]{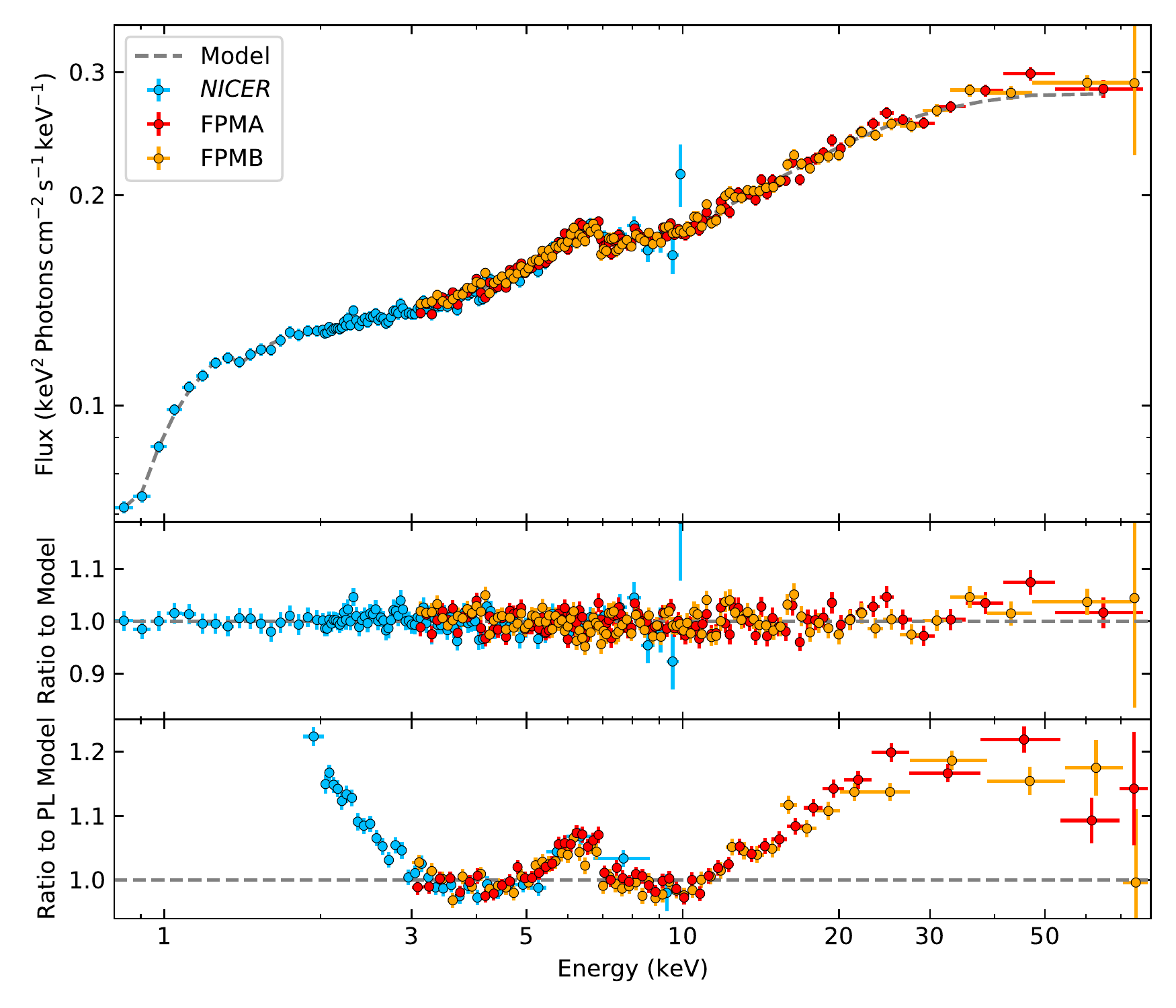}
\caption{\textit{Upper}: Unfolded spectrum of the joint \nustar and \nicer observations. The best-fit model for \nustar-FPMA is shown in the dashed grey line. To account for the cross-normalization terms (see Table~\ref{tab:august}), the FPMB and \nicer data are divided by 1.051 and 1.035, respectively.
\textit{Middle}: The ratio of data to the best-fit model.
\textit{Bottom}: The ratio of data to a simple power-law model. The model is fitted only to the 3--4\,keV and 10--12\,keV energy ranges.
The data have been rebinned for visual clarity. 
\label{fig:ausut_spec}
}
\end{figure*}

\subsection{Joint Analysis, 2020 August 16}  \label{subsec:spectral_joint2}

The upper panel of Figure~\ref{fig:ausut_spec} shows the \nustar 004 and simultaneous \nicer spectra, and the bottom panel presents the ratio of data to an absorbed power-law model (\texttt{tbabs*powerlaw}) fitted only to the 3--4\,keV and 10--12\,keV energy bands ($\Gamma\sim 1.8$). As reported by \citet{Yao2020_13957}, we clearly detected the broadened Fe K$\alpha$ line and Compton hump, characteristic of the relativistic reflection spectrum commonly seen in accreting X-ray binaries \citep{Garcia2011}. 

We modeled the spectrum by a combination of disk blackbody and relativistic reflection from an accretion disk (\texttt{tbfeo*edge*(simplcutx*diskbb+relxillCp)}, in \texttt{xspec}). In this model, the continuum is assumed to be produced by Comptonization of the disk photons (\texttt{simplcut*diskbb}, \citealt{Steiner2017, Mitsuda1984}), and the reflection is fitted with a \texttt{relxill} model \citep{Garcia2014, Dauser2014} that incorporates such continuum (\texttt{relxillCp}). A photoelectric absorption (\texttt{edge}) was added to account for instrumental uncertainties within the spectrum where \nicer's calibration is still ongoing (see, e.g., \citealt{Ludlam2020})

All data were fitted using $C$-statistics. For \nustar we fitted the data over the 3--79\,keV range, while for \nicer we fitted from 0.8 to 10\,keV. The \nustar data were grouped to have signal-to-noise ratio of 6 and oversampling factor of 3. Details of the model fitting are presented in Appendix~\ref{subsec:relxillCp}. The best-fit model is shown in Figure~\ref{fig:ausut_spec}. The model parameters are given in Table \ref{tab:august}.

The best-fit reflection spectrum is analogous to those observed in other black hole binaries, such as GX~339$-$4 \citep{Wang-Ji2018} or XTE~J1550$-$564 \citep{Connors2020}. 
% 1.4966e-09 --> 1.7633e-09
The unabsorbed flux in the 0.3--100\,keV band for FPMA is $1.8\times 10^{-9}{\rm erg\,cm^{-2}\,s^{-1}}$. At distances of [1, 3, 5, 10]\,kpc, this corresponds to a luminosity of [0.21, 1.9, 5.3, 21.1]$\times 10^{36}\,{\rm erg\,s^{-1}}$. Therefore, the X-ray luminosity on 2020 August 16 is $1.4\times 10^{-4} \lesssim L_{\rm X}/L_{\rm Edd} \lesssim 1.4\times 10^{-2}$ for a 10\,$M_\odot$ compact object.

\subsection{Joint Analysis, 2020 September 20}\label{subsec:spectral_joint3}

\begin{figure*}[htbp!]
\centering
\includegraphics[width=0.75\textwidth]{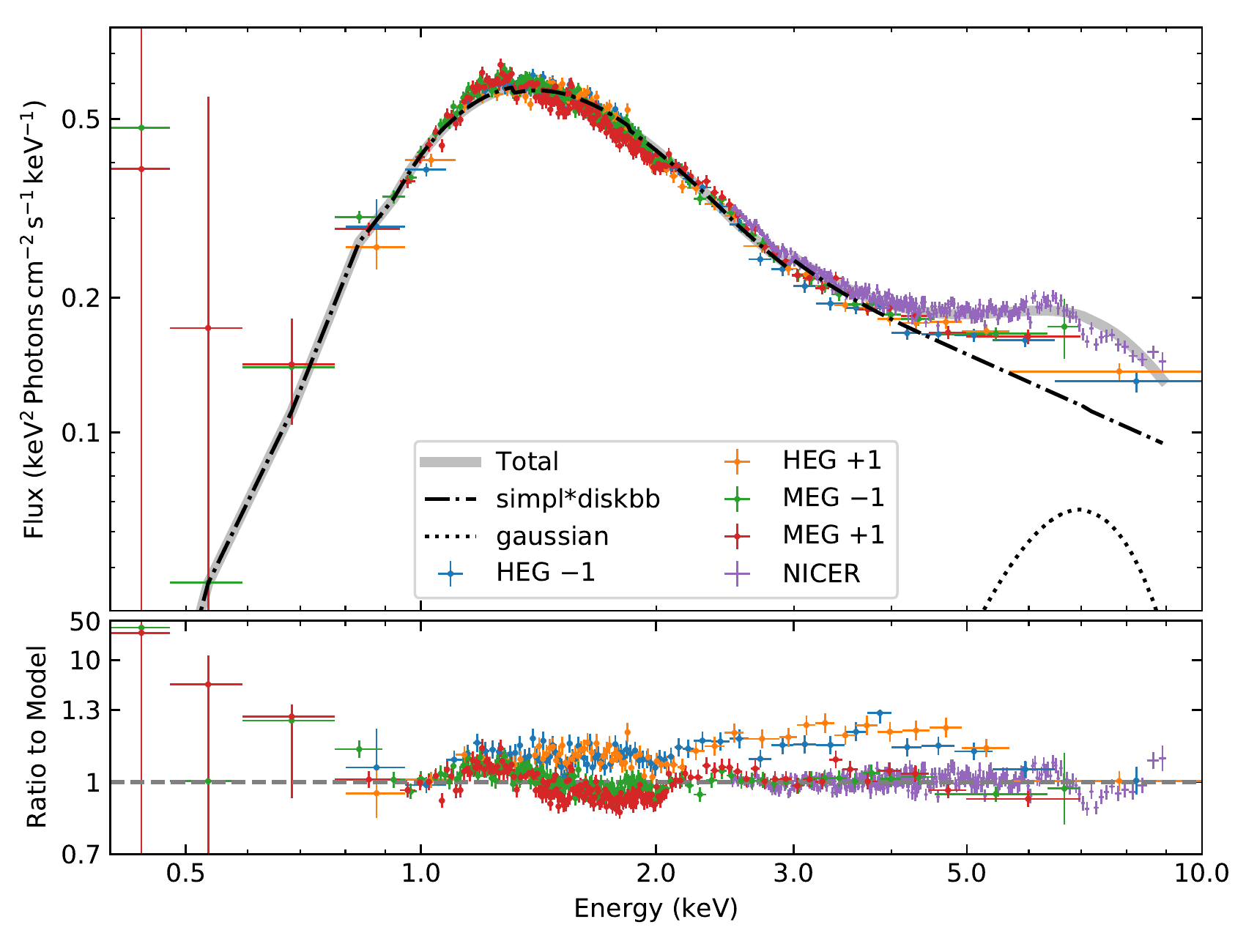}
\caption{\textit{Upper}: Unfolded spectrum of the joint \chandra and \nicer observations. We show the best-fit model for \nicer above 4.0\,keV, and the best-fit model for MEG $-1$ below 4.0\,keV. The total, \texttt{gaussian} component, and \texttt{simpl*diskbb} component are shown as the thick solid, thin dotted, and thin dash-dotted lines, respectively. To account for the cross-normalization terms (see Table~\ref{tab:sep}), the \nicer data are divided by 0.901.
\textit{Bottom}: The ratio of data to the best-fit model. The y-axis is shown in linear scale from 0.7 to 1.3, and in log scale from 1.3 to 50.
The data have been rebinned for visual clarity.
\label{fig:sep_spec}
}
\end{figure*}

\begin{deluxetable}{lr}[htbp!]
\tablecaption{Best-fit model parameters of the 2020 September 20 joint observations.  \label{tab:sep} }
\tablehead{
\colhead{Parameter \hspace{1.5cm} }  
& \colhead{ \hspace{1.5cm} 90\% Interval} }
\startdata
\multicolumn{2}{c}{\texttt{constant}} \\ \hline
$\mathcal{C_{\rm HETG}}$ & 1 (frozen)\\ % checked
$\mathcal{C_{\nicer}}$ & $ 0.901\pm 0.007$\\% checked
\multicolumn{2}{c}{\texttt{tbabs}} \\ \hline
$N_{\rm H}$ (10$^{22}$ \cmcm) & $0.417^{+0.014}_{-0.013}$ \\ %checked
\multicolumn{2}{c}{\texttt{simpl}} \\ \hline
$\Gamma$ & $ 2.80 \pm 0.05 $ \\ %checked
$f_{\rm sc}$ & $0.176 \pm 0.007$ \\
$R_{\rm up}$ & 1 (fixed) \\ 
\multicolumn{2}{c}{\texttt{diskbb}} \\ \hline
$T_{\rm disk}$ (keV) & $0.315 ^{+0.004}_{-0.005}$ \\% checked
R$_{\rm in}^{\ast}$\tablenotemark{$\dagger$} &  $136 \pm 6$ \\ % checked
\multicolumn{2}{c}{\texttt{gaussian}} \\ \hline
$E_{\rm line}$ (keV)  & 6.4 (fixed)  \\ 
$\sigma_{\rm line}$ (keV) &  $1.84\pm 0.09$ \\
Norm$_{\rm line}$\tablenotemark{$\ddagger$} & $0.0074\pm 0.0007$  \\
\hline
$\chi^2$ / \texttt{d.o.f.} & 12095.94 (23735) \\
\enddata
\tablenotetext{\dagger}{$R_{\rm in}^{\ast}$ has the same meaning as that in Table~\ref{tab:august}.}
\tablenotetext{\ddagger}{Normalization of the Gaussian in $\rm photon \,cm^{-2} \, s^{-1}$.}
\end{deluxetable}

The upper panel of Figure~\ref{fig:sep_spec} shows the simultaneous \chandra and \nicer observations. No strong narrow emission or absorption lines were detected in the HETGS spectrum. To model the continuum, we adopted the \texttt{constant*tbabs*(simpl*diskbb+gaussian)} model, where \texttt{simpl} is a Comptonization model that generates the power-law component via Compton scattering of a fraction ($f_{\rm sc}$) of input seed photons from the disk \citep{Steiner2009}. The flag $R_{\rm up}$ was set to 1 to only include upscattering. The \texttt{gaussian} component was added to account for the existence of a relativistic broadened iron line, and we fixed the line center ($E_{\rm line}$) at 6.4\,keV. We fitted the \nicer data over the 2.5--9.0 keV range. For HEG and MEG, we included the 0.8--10\,keV and 0.4--7.0\,keV bands, respectively. All data were fitted using $\chi^{2}$-statistics. 
The best-fit model is shown in Figure~\ref{fig:sep_spec}. The model parameters are given in Table \ref{tab:sep}.

\begin{figure*}[htbp!]
    \centering
    \includegraphics[width=\textwidth]{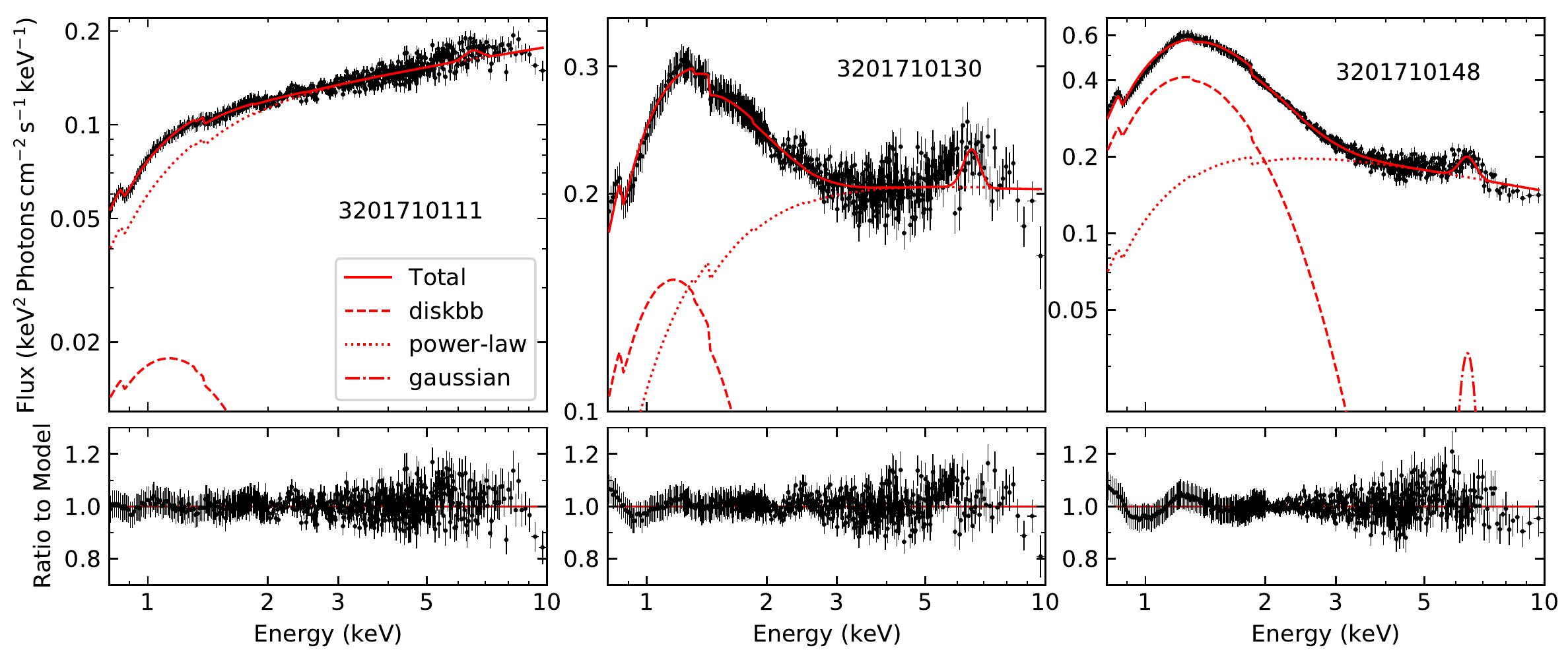}
    \caption{\nicer spectral fitting at three representative epochs, OBSID 3201710111, 3201710130, and 3201710148. The total flux in the best-fit model is shown as solid lines. The flux contribution from the \texttt{diskbb}, \texttt{pegpwrlw}, and \texttt{gaussian} components is shown as dashed, dotted, and dash-dotted lines, respectively. \label{fig:nicer_represent}}
\end{figure*}

\begin{figure}[htbp!]
    \centering
    \includegraphics[width=\columnwidth]{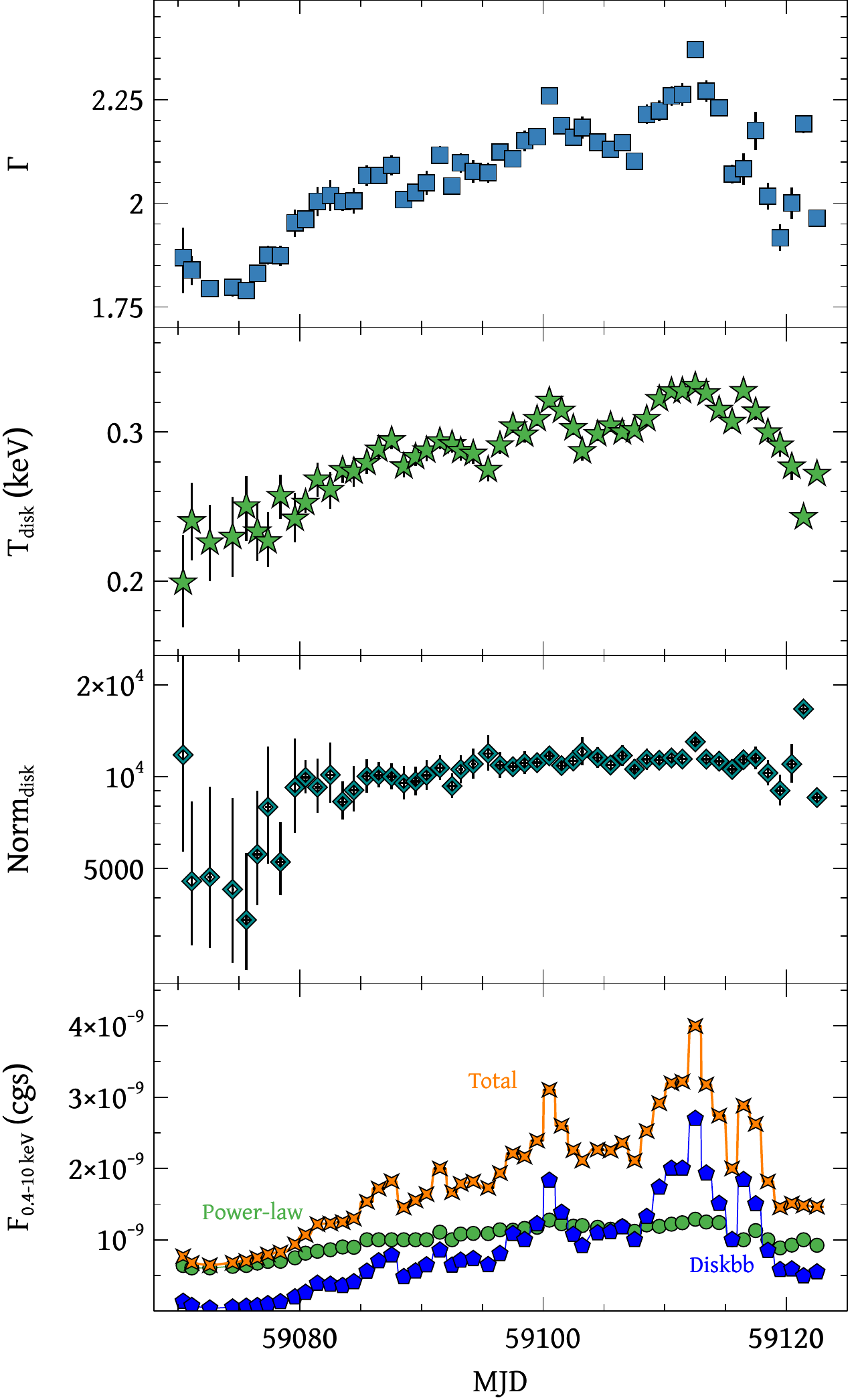}
    \caption{Evolution of \nicer spectral best-fit parameters, and unabsorbed 0.4--10\,keV flux from the \texttt{diskbb} component (blue pentagons), the \texttt{pegpwrlw} component (green circles), and the total (orange crosses).}
    \label{fig:nicer_spec_evolution}
\end{figure}

As can be seen from the bottom panel of Figure~\ref{fig:sep_spec}, the model underpredicts the MEG data below $\sim 0.8$\,keV. The MEG effective area below 1\,keV is sensitive to the correction for contamination, which currently undercorrects for the increasing depth of the contaminant. The magnitude of the effect is estimated to be about 20\% at 0.65\,keV and 10\% at 0.8\,keV, in the sense that estimated MEG fluxes should be even larger than shown in Figure~\ref{fig:sep_spec}.

The HETGS data can be used to constrain $N_{\rm H}$. By fitting a simple model to a limited wavelength range, the \ion{Mg}{I} and \ion{Ne}{I} edges due to the ISM can be determined directly. The continuum model in this case is empirical, a log-parabolic shape, and the edge is modeled in \texttt{isis} \citep{Houck2000} using the \texttt{edge} model, which has no structure at the edge but has the appropriate asymptotic behavior for the ISM edge. Fitting the 11--17\,\AA\ (0.73--1.13\,keV) region, we find that the \ion{Ne}{I} edge optical depth is 0.170$^{+0.06}_{-0.07}$, giving an estimate of $N_{\rm H}= 2.2^{+0.7}_{-0.9} \times 10^{21}\,{\rm cm^{-2}}$ \citep{Wilms2000}. An optical depth at the \ion{Ne}{I} edge of 0.33 is expected when $N_{\rm H} = 4.2 \times 10^{21}\,{\rm cm^{-2}}$, which is ruled out at the 4.3$\sigma$ level. An independent measurement from fitting the \ion{Mg}{I} line in the 8--11\,\AA\  (1.13--1.55\,keV) region gives an optical depth of $0.043^{+0.021}_{-0.014}$, and $N_{\rm H} =3.1^{+1.5}_{-1.0} \times 10^{21}\,{\rm cm^{-2}}$. 

Paper II measured the equivalent width ($EW$) of \ion{Na}{I} D line and diffuse interstellar bands (DIBs) from a summed optical spectrum, and constrained the line-of-sight extinction to be $0.8\lesssim E(B-V) \lesssim 1.2$. Using the calibration of $N_{\rm H} = 5.55\times 10^{21}\times E(B-V)$ \citep{Predehl1995}, the line-of-sight column density can be inferred to be $4.4 < N_{\rm H}/(10^{21}\,{\rm cm^{-2}}) < 6.7$. This is consistent with the $N_{\rm H}$ derived from the continuum fit. Therefore, in the \nicer-only spectral analysis (Section~\ref{subsec:spectral_nicer}), we adopt $N_{\rm H} = 5\times 10^{21}\,{\rm cm^{-2}}$.

\subsection{\nicer-only Spectral Analysis} \label{subsec:spectral_nicer}
 
Figure~\ref{fig:nicer_represent} shows the \nicer spectra at three representative epochs (59076, 59100, and 59113\,MJD). The continuum can be described by a combination of a multi-color disk component, a power-law component, and a Gaussian line component at 6.3--6.5\,keV. To investigate the evolution of spectral components, for each OBSID, we fitted a \texttt{tbfeo*(diskbb+pegpwrlw+gaussian)*edge} model to the 0.8--10\,keV \nicer spectrum. The \texttt{edge} feature at $\approx$1.4\,keV was included, as found to be present in the \nicer and \nustar joint spectral analysis (Section~\ref{subsec:spectral_joint2}). In the \texttt{tbfeo} model, the O and Fe abundances were fixed at Solar values, and $N_{\rm H}$ was fixed at $5\times10^{21}\,{\rm cm^{-2}}$. All data were fitted using $\chi^{2}$-statistics. The best-fit models provided a reduced-$\chi^2$ close to 1 in most of the cases. 
%the line width is not well constrained due to the weak nature of the line in \nicer data.  

The evolution of spectral parameters of the hydrogen column density $N_{\rm H}$, the power-law photon index $\Gamma$, the temperature at inner disk radius $T_{\rm disk}$, and the disk-blackbody normalization term ${\rm Norm}_{\rm disk}=(R_{\rm in}/D_{10})^2{\rm cos}i$ are shown in Figure~\ref{fig:nicer_spec_evolution}. ${\rm Norm}_{\rm disk}$ remained almost constant after 59082\,MJD. This provides evidence that the inner disk radius ($R_{\rm in}$) remained at $\sim100$--1000\,km assuming a range of distances from $D\sim 10$\,kpc to 1\,kpc. 

In the bottom panel of Figure~\ref{fig:nicer_spec_evolution}, we present the unabsorbed 0.4--10\,keV fluxes in the disk-blackbody component, the power-law component, and the total (disk-blackbody $+$ power-law $+$ Gaussian). Note that the Fe line flux is significantly smaller than the other two components. The occasional enhancement observed in the source light curve matches to the brightening of the thermal component.

\section{Discussion} \label{sec:discussion}

\subsection{X-ray States in AT2019wey}
The X-ray spectral-timing properties of \target are in line with the typical properties of LMXBs in the LHS and HIMS. 

From 2019 December to 59082\,MJD (2020 August 21), \target stayed in the canonical LHS of LMXBs. In the first six months, the spectrum was dominated by a hard power-law component ($1.7 \lesssim \Gamma \lesssim 2.0$) with little contribution from the disk component (Figures~\ref{fig:april_spec} and \ref{fig:nicer_spec_evolution}). It moved along the HL on the RID (Figure~\ref{fig:nicer_rid}), and the fractional rms stayed at $\sim$30\%. No QPO was observed (Section~\ref{subsec:timing}). The X-ray color softened as the source brightened (Figure~\ref{fig:nicer_hid}). Toward the end of the brightening, spectral features of relativistic reflection were clearly seen (Figure~\ref{fig:ausut_spec}). Modeling of the reflection spectrum suggests a small inclination ($i\lesssim 30^{\circ}$, see Appendix~\ref{subsec:relxillCp}). The rms variability decreased with increasing photon energy (right panel of Figure~\ref{fig:rms_E}), indicating that cooler regions of the source are more variable than hotter regions, perhaps due to inhomogeneities in an accretion disk.

Between 59082\,MJD and 59122\,MJD (2020 September 30), \target was in the canonical HIMS of LMXBs. The power-law component steepened ($2.0\lesssim \Gamma \lesssim 2.3$), and the thermal disk emission became comparable to the power-law component in the 0.4--10\,keV band (Figure~\ref{fig:nicer_spec_evolution}). The excess in the very soft X-ray band (Section~\ref{subsec:spectral_joint3}) might arise from reprocessing of X-rays in the outer accretion disk. Its soft X-ray light curve underwent a few episodes of mini-outbursts, which were correlated with the enhancement of a thermal component. At the same time, the source left the HL on the RID as the fractional rms decreased (Figure~\ref{fig:nicer_rid}). A weak type-C LFQPO was observed, and its characteristic frequency increased from $\sim$2\,Hz to $\sim$6.5\,Hz as the disk flux increased. It did not reach the SIMS since the fractional rms was $>9$\% at the minimum (Figure~\ref{fig:nicer_hrd}).

\target likely stayed in the HIMS from 2020 October 1 to November 30, since the 2--10\,keV and 15--50\,keV light curves remained roughly constant (Figure~\ref{fig:xlc}). We note that after being active in X-ray for at least $\sim12$\,months, \target had not transitioned to the SIMS or HSS. The lack of hysteresis in the HID (Figure~\ref{fig:nicer_hid}) is similar to the BH candidate MAXI\,J1836$-$194 \citep{Russell2013}.

Paper II reported the radio brightening as AT2019wey transitioned from LHS to HIMS. \citet{Yadlapalli2021} reported the detection of a resolved radio source during the HIMS, which was interpreted as a steady compact jet. The evolution of the radio emission is consistent with LMXBs in the hard states \citep{Fender2004, Migliari2006}. 

The X-ray properties observed in \target thus far make it a promising candidate for the population of ``hard-only'' outbursts \citep{Tetarenko2016}. The distance of this system is poorly constrained to $\sim$1--10\,kpc (Paper II). Given the brightness of \target in the optical ($r\approx 17.4$\,mag), the $Gaia$ mission will be able to determine the parallax to the source and thus settle the distance. Assuming a typical distance at 3--5\,kpc, the 0.3--100\,keV X-ray luminosity of \target remained at a few times $10^{35}\,{\rm erg\,s^{-1}}$ (Section~\ref{subsec:spectral_joint1}) for $\sim6$\,months in the LHS, increased by a factor of $\sim10$ to a few times $10^{36}\,{\rm erg\,s^{-1}}$ (Section~\ref{subsec:spectral_joint2}) over $\sim2$\,months, and stayed at this luminosity afterward in the HIMS. This range of X-ray luminosities is at the lower end of the whole population of BH transients, but is typical for ``hard-only'' outbursts \citep{Tetarenko2016}. 

\begin{deluxetable*}{lccccc}[htbp!]
	\tablecaption{Short-period ($P_{\rm orb}<16$\,hr) BH or BH candidate LMXB outbursts discovered from 2009 to 2020. \label{tab:shortP_bhxb}}
	\tablehead{
		\colhead{Name} &
		\colhead{$P_{\rm orb}$ (hr)} &
		\colhead{Discovery Instrument} &
		\colhead{Discovery Date}&
		\colhead{X-ray States} & 
		\colhead{References} 
	}
	\startdata
	\target & $<16$ & ATLAS; \srg  & 2019 Dec 7 
	& LHS, HIMS & 1, 2, 23, 24 \\
	\hline
	MAXI\,J1305$-$704 & 9.7 & \maxi & 2012 Apr 9 & IMS & 3, 4, 5, 25\\
	\hline
	\multirow{3}{*}{Swift\,J1357.2$-$0933} & \multirow{3}{*}{2.8} &  \swift/BAT & 2011 Jan 28 & LHS & 6, 7, 8\\
	& & CRTS & 2017 Apr 20 & LHS & 9, 10 \\
	& & ZTF & 2019 Mar 31 & -- & 11 \\
	\hline
	MAXI\,1659$-$152 & 2.4 & \swift/BAT, \maxi  & 2010 Sep 25 & LHS, IMS, HSS & 5, 12, 13, 14  \\ 
	\hline
	IGR\,J17451$-$3022 & 6.3 & \integral & 2014 Aug 22--24 & HSS & 20, 21, 22 \\
	\hline
	XTE\,J1752$-$223 & $<6.8$ &  \rxte & 2009 Oct 23 & LHS, IMS, HSS & 5, 15, 16 \\
	\hline
	MAXI\,1836$-$194 & $<4.9$ & \maxi, \swift/BAT & 2011 Aug 30 & LHS, HIMS & 17, 18, 19 \\
	\enddata
	\tablecomments{
	Instruments: the International Gamma-Ray Astrophysics Laboratory (\integral; \citealt{Winkler2003}); the Rossi X-ray Timing Explorer (\rxte; \citealt{Swank1999}); the Catalina Real-Time Transient Survey (CRTS; \citealt{Drake2009}); the Zwicky Transient Facility (ZTF; \citealt{Bellm2019b}; \citealt{Graham2019}).
	References. (1) This work (2) Paper II 
	(3) \citet{Sato2012} (4) \citet{Shidatsu2013} (5) \citet{Tetarenko2016} (6) \citet{Krimm2011} (7) \citet{Corral-Santana2013} (8) \citet{Armas-Padilla2013} (9) \citet{Drake2017} (10) \citet{Beri2019} (11) \citet{vanVelzen2019} (12) \citet{Negoro2010} (13) \citet{Mangano2010} (14) \citet{Kuulkers2013} (15) \citet{Markwardt2009} (16) \citet{Ratti2012} (17) \citet{Negoro2011} (18) \citet{Ferrigno2012} (19) \citet{Russell2014} (20) \citet{Chenevez2014} (21) \citet{Jaisawal2015} (22) \citet{Bozzo2016} (23) \citet{Tonry2019} (24) \citet{Mereminskiy2020} (25) \citet{Morihana2013} }
\end{deluxetable*}

\begin{figure*}
    \centering
    \includegraphics[width=0.93\textwidth]{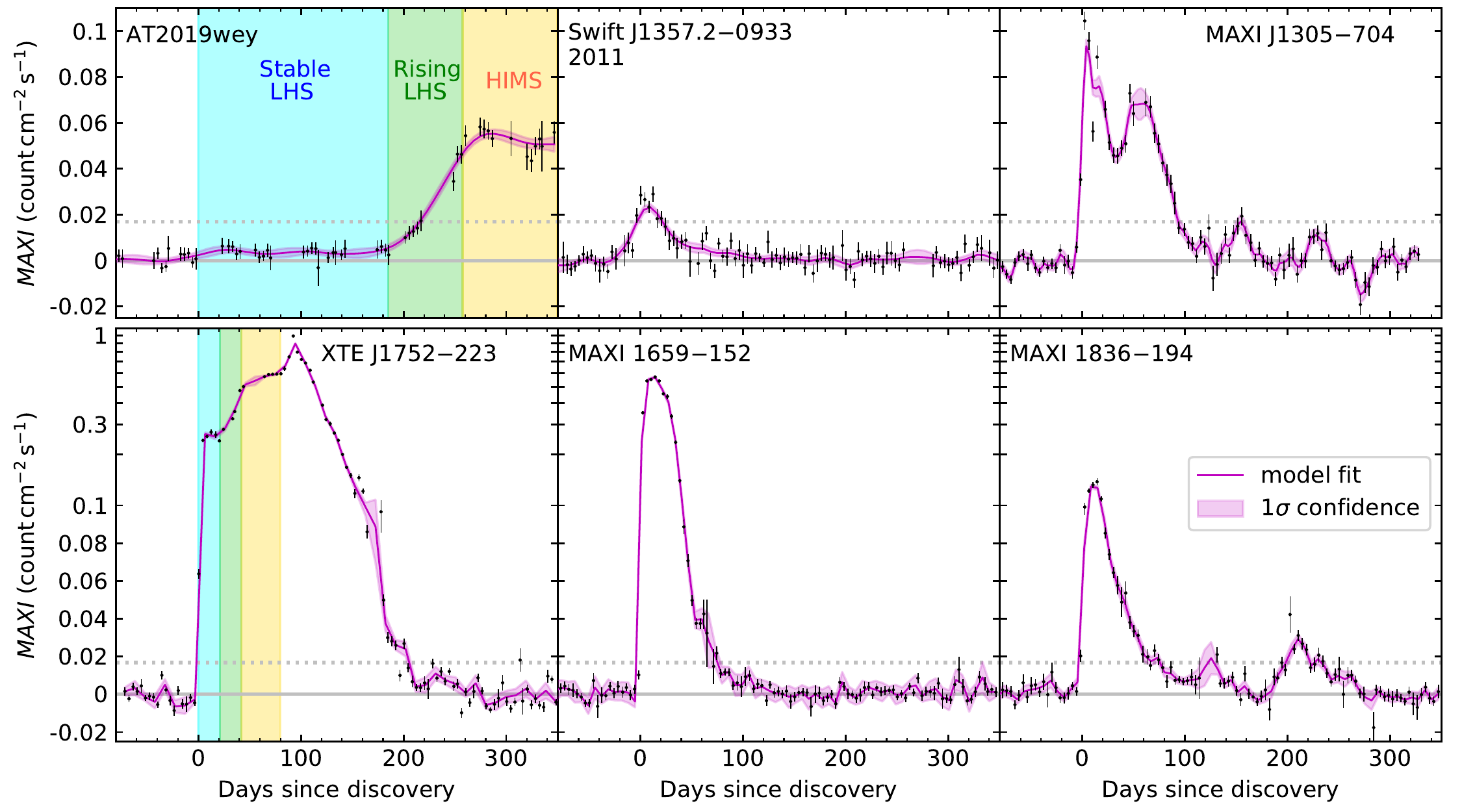}
    \caption{\maxi 2--10\,keV Light curves of outbursts from short-period systems. The background of \target and XTE\,J1752$-$223 are color-coded to emphasize the two stable flux levels during the LHS and HIMS. 
    In each panel, the black data points are 4-day binned light curves. The magenta curve is a model fit to the data, generated with a Gaussian process following procedures outlined in Appendix B.4 of \citet{Yao2020dge}. The dotted horizontal line marks \maxi 4\,day 3$\sigma$ detection limit of 7\,mCrab in the 3--10 keV band \citep{Negoro2016}. The y-axis is shown in linear scale below 0.1, and in log scale above 0.1. The light curves of MAXI\,J1305$-$704 and XTE\,J1752$-$223 are affected by leak counts from nearby bright sources periodically due to the 72\,days 
    ISS orbital precessions. We have subtracted the average count rate between 2015 and 2020 to remove first-order contamination. 
    \label{fig:outbursts}}
\end{figure*}

\subsection{Nature of the Compact Object}
NS signatures of coherent pulsations and thermonuclear X-ray bursts were not detected in 394\,ks of \nicer and 80\,ks of \nustar data (see Appendix~\ref{sec:details}). The X-ray spectral and timing properties shown in this paper are consistent with both NS and BH LMXB outbursts. However, a few properties of this source favor a BH accretor.

First of all, during the initial six months of the LHS, the power-law index was $\Gamma\approx 1.77$ and the 0.5--10\,keV luminosity was $4.5\times 10^{33}$--$4.5\times 10^{35}\,{\rm erg\,s^{-1}}$ (Section~\ref{subsec:spectral_joint1}). 
This makes AT2019wey closer to BH binaries on the $\Gamma$--$L_{\rm X}$ diagram (see Fig.~2 of \citealt{Wijnands2015}).
Moreover, the positions of this source on the $L_{\rm radio}$--$L_{\rm X}$ and the $L_{\rm opt}$--$L_{\rm X}$ diagrams are also closer to BH binaries (Paper II).

Therefore, although we can not preclude the possibility of a NS at this time, it is highly suggestive that AT2019wey is a BH system.

\subsection{The Slow Rise of the Outburst}

LMXB outbursts (also termed as X-ray novae) span a wide range of morphological types \citep{Chen1997}. Theories for the canonical fast-rise exponential-decay (FRED) profile of X-ray novae have been developed based on the disk instability model (DIM), which was originally invoked to explain dwarf nova outbursts \citep{Lasota2001}. Disk truncation and irradiation are generally invoked to account for the longer evolution timescale and recurrence time of X-ray novae \citep{vanParadijs1996, Dubus2001}. Recently, detailed analysis of the decay profile of X-ray outbursts provides evidence for the existence of generic outflows and time-varying irradiation \citep{Tetarenko2018Nat, Tetarenko2018, Shaw2019, Tetarenko2020}.

Here we focus on the rise profile of AT2019wey. Paper II shows that the orbital period of AT2019wey is likely less than 16\,hours. To compare \target with other short-period LMXBs, we select outbursts discovered between 2009 and 2020 from the BlackCAT\footnote{\url{http://www.astro.puc.cl/BlackCAT/index.php}} catalog \citep{Corral-Santana2016}. Systems with $P_{\rm orb}\lesssim 16$\,hours are summarized in Table~\ref{tab:shortP_bhxb}. Figure~\ref{fig:outbursts} shows their \maxi 2--10\,keV light curves. We excluded IGR\,J17451$-3022$ since its \maxi data was highly contaminated by the bright persistent source 1A\,1742$-$294. We also excluded the 2017 and 2019 outbursts of Swift\,J1357.2$-$0933 since their X-ray fluxes were too faint to be seen by \maxi --- they were only detected by follow-up observations conducted by \nustar, \swift/XRT, and \nicer \citep{Beri2019, Beri2019b, Gandhi2019, Rao2019}. 

Figure~\ref{fig:outbursts} (middle and right panels) show that the 2--10\,keV light curves of MAXI\,J1305$-$704, MAXI\,J1659$-$152, MAXI\,J1836$-$194, and the 2011 outburst of Swift\,J1357.2$-$0933 rose to maximum in 5--20\,days. In comparison, the evolution of \target's light curve (upper left panel of Figure~\ref{fig:outbursts}) is rather slow. Its 2--10\,keV flux rose to $\sim1$\,mCrab upon discovery, remained at this level for about 6 months, and brightened to a maximum of only $\sim20$\,mCrab afterwards. This is similar to the initial evolution of XTE\,J1752$-$223 (lower left panel of Figure~\ref{fig:outbursts}), where the source stayed in the hard state with two stable flux levels for about 3 months \citep{Nakahira2010}\footnote{The exact time of the LHS $\rightarrow$ HIMS transition was not well determined for XTE\,J1752$-$223 \citep{Brocksopp2013}.}. In the left panels of Figure~\ref{fig:outbursts}, we color-code the background of the two stable flux levels by blue and yellow, and the rising between the two stable levels by green. As mentioned by \citet{Nakahira2010}, the long duration of the initial LHS and the two plateau phases are rather uncommon for recorded LMXB outbursts, and might be accounted for by a slow increase of $\dot M$. We note that XTE\,J1752$-$223 later transitioned to the HSS and completed the hysteresis pattern on the HID. It remains to be seen if AT2019wey will transition to the HSS.
% New recipes in the DIM are perhaps needed to explain such an intriguing behaviour.

\section{Conclusion} \label{sec:conclusion}

In this paper, we present \nicer, \nustar, \chandra, \swift, and \maxi observations of the X-ray transient AT2019wey (SRGA\,J043520.9$+$552226, SRGE\,J043523.3$+$552234). By analyzing its spectral-timing properties, we conclude that \target is a LMXB outburst with a BH or NS accretor. The source's evolution from 2019 December to 2020 November can be separated by three phases: the stable LHS from 2019 December to 2020 May ($\sim1$\,mCrab), the rising LHS from 2020 June to August, and the stable HIMS from 2020 August to November ($\sim20$\,mCrab). 

The long duration of the initial LHS and the two plateau phases of AT2019wey (Figure~\ref{fig:outbursts}) are not commonly seen. We searched the literature for analogs of AT2019wey. The closest analog we found is XTE\,J1752$-$223, a candidate BH LMXB with an orbital period of $<7$\,hr (Table~\ref{tab:shortP_bhxb}). 

If \srg had not discovered AT2019wey in 2020 March, the source would have probably been discovered by \maxi or BAT during the HIMS, and in a retrospective fashion, the initial $\sim1$\,mCrab flux excess could have been revealed by \maxi long-term monitoring. However, the \srg discovery is important to trigger rapid X-ray follow-up observations, which classify the initial plateau phase as in the LHS. 

The repeated \srg all-sky surveys that are being carried offer the opportunity to discover other events similar to \target at early epochs (and thus enable critical multi-wavelength follow-up).
Furthermore, the  eROSITA sensitivity is unprecedented: $<5\times 10^{-14}\,{\rm erg\,s^{-1}\,cm^{-2}}$ (1.3\,$\mu$Crab) in the 0.3--2.2\,keV band, and $<7\times 10^{-13}\,{\rm erg\,s^{-1}\,cm^{-2}}$ (36\,$\mu$Crab) in the 2.3--8\,keV band \citep{Predehl2021}. This sensitivity should lead to the discovery of fainter versions of AT2019wey look-alikes.

%%%%% 

\begin{acknowledgements}
We thank the anonymous reviewer for providing comments that have largely improved this manuscript. We thank Belinda Wilkes and Patrick Slane for allocating DD time on \chandra, and Fiona Harrison for allocating \nustar DD time. We thank Kumiko Morihana for providing the cleaned \maxi light curve of MAXI\,J1305$-$704. Y.Y. thanks the Heising–Simons Foundation for support. R.M.L. acknowledges the support of NASA through Hubble Fellowship Program grant HST-HF2-51440.001. J.A.G. acknowledges support from NASA grant 80NSSC17K0515 and from the Alexander von Humboldt Foundation. Z.W. acknowledges support from the NASA postdoctoral program. 
\nicer research at NRL is supported by NASA.
This work was partially supported under NASA contract No. NNG08FD60C and made use of data from the \nustar mission, a project led by the California Institute of Technology, managed by the Jet Propulsion Laboratory, and funded by the National Aeronautics and Space Administration. We thank the \nustar Operations, Software, and Calibration teams for support with the execution and analysis of these observations. This research has made use of the \nustar Data Analysis Software (\texttt{nustardas}), jointly developed by the ASI Science Data Center (ASDC, Italy) and the California Institute of Technology (USA).
This work made use of data supplied by the UK \swift Science Data Centre at the
University of Leicester.

\end{acknowledgements}

\software{
\texttt{astropy} \citep{Astropy2013},
\texttt{CIAO} \citep{Fruscione2006},
\texttt{HEASoft} (v6.27; \citealt{HEASARC2014}),
\texttt{HENDRICS} \citep{Bachetti2015_MaLTPyNT}, 
\texttt{isis} \citep{Houck2000},
\texttt{matplotlib} \citep{Hunter2007},
\texttt{pandas} \citep{McKinney2010},
\texttt{PRESTO} \citep{Ransom2011};
\texttt{relxill} (v1.3.10; \citealt{Garcia2014, Dauser2014})
\texttt{Stingray} \citep{Huppenkothen2019},
\texttt{xspec} (v12.11.0; \citealt{Arnaud1996})
}

\facilities{NICER, NuSTAR, CXO (ACIS-S), Swift (XRT, BAT), MAXI}

\appendix
\section{Observation Logs}
Here we present observation logs of \nicer (Table~\ref{tab:nicer}), \nustar (Table~\ref{tab:nustar}), and \swift/XRT (Table~\ref{tab:XRT}).

\begin{deluxetable*}{ccc|ccc|ccc}[htbp!]
\tablecaption{\nicer Observation Log \label{tab:nicer}}
\tablehead{
\colhead{OBSID}  
& \colhead{Exp.} 
& \colhead{Start Time} 
& \colhead{OBSID}  
& \colhead{Exp.} 
& \colhead{Start Time} 
& \colhead{OBSID}  
& \colhead{Exp.} 
& \colhead{Start Time} \\
\colhead{}
& \colhead{(ks)} 
& \colhead{(UT)} 
&\colhead{}
& \colhead{(ks)} 
& \colhead{(UT)} 
&\colhead{}
& \colhead{(ks)} 
& \colhead{(UT)} 
}
\startdata
3201710105 & 0.22 & 2020-08-09 08:28 &
3201710106 & 1.10 & 2020-08-10 01:30 &
3201710107 & 4.82 & 2020-08-11 09:59 \\
3201710108 & 6.02 & 2020-08-12 07:43 &
3201710109 & 4.39 & 2020-08-13 08:30 &
3201710110 & 8.00 & 2020-08-14 04:38 \\
3201710111 & 10.55 & 2020-08-15 00:54 &
3201710112 & 4.21 & 2020-08-16 03:01 &
3201710113 & 4.85 & 2020-08-17 00:43 \\
3201710114 & 2.52 & 2020-08-18 13:52 &
3201710115 & 7.41 & 2020-08-19 02:14 &
3201710116 & 3.15 & 2020-08-20 04:51 \\
3201710117 & 2.32 & 2020-08-21 02:37 &
3201710118 & 6.39 & 2020-08-22 02:07 &
3201710119 & 3.57 & 2020-08-23 00:59 \\
3201710120 & 5.87 & 2020-08-24 00:25 &
3201710121 & 8.95 & 2020-08-24 23:36 &
3201710122 & 7.58 & 2020-08-26 03:29 \\
3201710123 & 7.76 & 2020-08-27 02:43 &
3201710124 & 10.07 & 2020-08-28 00:25 &
3201710125 & 3.65 & 2020-08-29 01:12 \\
3201710126 & 6.54 & 2020-08-30 00:26 &
3201710127 & 7.99 & 2020-08-31 02:49 &
3201710128 & 4.84 & 2020-09-01 00:31 \\
3201710129 & 2.97 & 2020-09-02 01:18 &
3201710130 & 3.71 & 2020-09-03 00:31 &
3201710131 & 6.90 & 2020-09-04 01:18 \\
3201710132 & 8.44 & 2020-09-05 00:32 &
3201710133 & 5.43 & 2020-09-06 02:46 &
3201710134 & 12.36 & 2020-09-07 00:05 \\
3201710135 & 13.69 & 2020-09-08 01:08 &
3201710136 & 23.53 & 2020-09-09 00:29 &
3201710137 & 14.38 & 2020-09-10 01:13 \\
3201710138 & 5.67 & 2020-09-11 01:47 &
3201710139 & 8.76 & 2020-09-11 23:34 &
3201710140 & 11.32 & 2020-09-13 03:34 \\
3201710141 & 10.02 & 2020-09-14 02:35 &
3201710142 & 10.49 & 2020-09-15 02:07 &
3201710143 & 5.80 & 2020-09-16 02:51 \\
3201710144 & 5.45 & 2020-09-17 03:40 &
3201710145 & 8.38 & 2020-09-18 02:58 &
3201710146 & 10.22 & 2020-09-19 00:12 \\
3201710147 & 16.59 & 2020-09-20 01:03 &
3201710148 & 6.54 & 2020-09-21 00:13 &
3201710149 & 11.19 & 2020-09-21 23:27 \\
3201710150 & 8.84 & 2020-09-23 03:24 &
3201710151 & 5.97 & 2020-09-24 01:02 &
3201710152 & 3.34 & 2020-09-25 01:52 \\
3201710153 & 3.82 & 2020-09-26 01:06 &
3201710154 & 2.53 & 2020-09-27 01:52 &
3201710155 & 3.11 & 2020-09-28 01:06 \\
3201710156 & 4.25 & 2020-09-29 00:20 &
3201710157 & 6.22 & 2020-09-30 01:10 &
&&\\
\enddata
\end{deluxetable*}

\begin{deluxetable}{cccc}[htbp!]
\tablecaption{\nustar Observation Log \label{tab:nustar}}
\tablehead{
\colhead{OBSID}  
& \colhead{Exp.} 
& \colhead{Start Time} 
& \colhead{Count Rate} \\
\colhead{}
& \colhead{(ks)}
& \colhead{(UT)}
& \colhead{(\cps)}
}
\startdata
90601315002 & 38 &  2020-04-18 11:21 & $2.3\pm0.7$ \\
90601315004 & 42 &  2020-08-16 12:16 & $30.8\pm2.6$ \\
90601315006 & 37 &  2020-08-27 02:51 & $35.1\pm2.7$ \\
\enddata
\end{deluxetable}

\begin{deluxetable}{ccccch}[htbp!]
\tablecaption{\swift/XRT Observation Log\label{tab:XRT}}
\tablehead{
\colhead{OBSID}  
& \colhead{Exp.} 
& \colhead{Start Time} 
& \colhead{Mode}
& \colhead{ Count Rate} 
%& \colhead{Unabsorbed Flux}
\\
\colhead{}
& \colhead{(s) }
& \colhead{(UT) }
& \colhead{}
& \colhead{(\cps)}
%& \colhead{($10^{-11}$\,erg\,cm$^{-2}$\,s$^{-1}$ )}
}
\startdata
13313001 & 1523 & 2020-04-12 06:07  & PC &  $0.645 \pm 0.029$  &  $4.84 \pm 0.22$ \\
13313002 & 874 & 2020-04-17 19:55  & PC &  $0.570 \pm 0.035$  &  $4.27 \pm 0.26$ \\
13313003 & 1026 & 2020-04-24 14:28  & PC &  $0.639 \pm 0.036$  &  $4.79 \pm 0.27$ \\
13313004 & 1043 & 2020-04-28 13:56  & PC &  $0.717 \pm 0.051$  &  $5.38 \pm 0.38$ \\
13313010 & 434 & 2020-09-02 20:36  & WT &  $27.57^{+0.28}_{-0.31}$  &  $325.9^{+12.5}_{-12.0}$ \\
13313011 & 1023 & 2020-09-09 16:40  & WT &  $42.50^{+1.58}_{-1.54}$  &  $660.8^{+13.3}_{-13.1}$ \\
13313012 & 858 & 2020-09-16 16:01  & WT &  $43.32^{+0.26}_{-0.29}$  &  $652.3^{+56.0}_{-50.8}$ \\
13313013 & 794 & 2020-09-23 20:03  & WT &  $40.53^{+2.35}_{-2.27}$  &  $789.2^{+18.6}_{-18.2}$ \\
\enddata
\tablecomments{Count rate is given in the 0.3--10\,keV band.}
\end{deluxetable}

\section{Details of Analysis}  \label{sec:details}

\subsection{\nicer Pulsation Search}\label{subsec:nicer_pulsations}
Pulsation searches were carried out for all \nicer data presented in Table~\ref{tab:nicer}. The \nicer data contains $2257$ GTIs spread over $394$\,ks of observations. Upon cursory inspection of the data with {\tt NICERsoft}\footnote{\url{https://github.com/paulray/NICERsoft}}, we found that detectors 34 and 43 suffered from high optical loading. Thus, the events in these detectors were excluded. The events were barycentered 
%using ${\rm RA}=68.84698^\circ$, ${\rm DEC}=+55.37619^\circ$ (equinox J2000.0), and with the JPL-DE405 solar system ephemeris 
using \texttt{barycorr}. We employed acceleration search and stacked power spectral search schemes to search for pulsations. 

To start with, we searched for pulsations using acceleration search. To account for possible frequency shifts due to binary Doppler motion, we employed an acceleration search algorithm over the $f$-$\dot{f}$ plane in the PulsaR Exploration and Search TOolkit (\texttt{PRESTO}\footnote{\url{https://www.cv.nrao.edu/~sransom/presto/}}; \citealt{Ransom2011}). The acceleration search is valid under the assumption that the pulsar has a constant acceleration throughout the observation, and is most effective for observation durations of $T \lesssim P_{\rm orb}/10$ \citep{Ransom2002}. 

To determine the GTIs (and hence event files) used in the acceleration searches, we started from the $2257$ GTIs in the original filtered event file. In order to prevent very short GTIs from being used, adjacent GTIs that were less than 11\,s apart were combined. This resulted in a total of $445$ GTIs, ranging in length from 1\,s to 2648\,s. We imposed a minimum GTI duration of 64\,s to avoid spurious signals in short GTIs, leaving 378 GTIs with a median length of 883\,s. For each of these GTIs (considered independently), we further filtered events from three energy ranges: 0.5--2\,keV, 2--12\,keV, and 0.5--12\,keV. The 1134 event files were then extracted with \texttt{niextract-events}. We then ran the search using the \texttt{accelsearch} task in \texttt{PRESTO} over the range 1--1000\,Hz, positing that Doppler shifting would cause the possible signal to drift across a maximum of 100 Fourier frequency bins. For the median GTI length (883\,s) and a fiducial fundamental pulsation frequency of 300\,Hz, this corresponds to accelerations of up to $a = z_{\rm max}c/(fT^2) \approx 130{\rm\, m\,s^{-2}}$. The typical acceleration in a NS LXMB, say in a $12$-hour orbit around a $0.2\,M_\odot$ companion, is approximately $5.7{\rm\,m\,s^{-2}}$. The acceleration searches yielded no candidate signals above the statistical significance threshold of 3-$\sigma$, after accounting for the total number of trials. 

An alternative pulsation search algorithm involves stacking power spectra from $M$ segments and calculating an averaged power spectrum. This is Bartlett's method \citep{Bartlett1948}, in which the original time series is broken up into $M$ non-overlapping segments of equal length. 
The $M$ segments were binned at $\Delta t = 0.5{\rm\,ms}$, such that we sampled at the Nyquist frequency of $1000{\rm\, Hz}$. 
The Leahy-normalized power spectrum was then computed for each of the $M$ segments, using the \texttt{realfft} task in \texttt{PRESTO} \citep{Leahy1983}. 
Finally, the $M$ resulting spectra were averaged and the corresponding noise distributions were calculated. 
The detection level for any candidate signal was then determined by calculating the probability that the power in any frequency bin exceeded that of a detection threshold (say, 3-$\sigma$). 
This was calculated through the integrated probability of the $\chi^2$ distribution with $2MW$ degrees of freedom, with $W$ being the rebinning factor \citep{vanderKlis1988}. The stacking procedure was done to enhance the signal of faint millisecond pulsars. %It reduced the variance of the noise-induced fluctuations, at the expense of coarser frequency resolution.

The stacked power spectra were calculated with segments of length 64, 128, 256, and 512\,s, to account for possible orbital modulations in the pulsar frequency with yet unknown binary parameters. 
On top of stacking the power spectra from segments of the entire time series, the stacked power spectra were also calculated for various sub-time series, where the choices were informed by the overall light curve binned at 128\,s and looking at the source brightness level. 
The number of segments admitted into the calculation for the stacked power spectrum also depends on a segment threshold (in \%).
That is, for each segment, a 1-s binned light curve was generated. If the fraction of bins with counts is less than the threshold, then that segment will not be used in the calculation. 
Segment thresholds used were 20\%, 50\%, 70\%, and 100\%. 
We also searched over energy ranges 0.5--2\,keV, 2--12\,keV, and 0.5--12\,keV. 
The averaged power spectrum was finally calculated by dividing the total power spectrum by the number of segments used.

From all of these stacked power spectra, there were no candidate signals that exceeded the 3-$\sigma$ detection level, after accounting for the total number of trials.

\subsection{\nustar Pulsation Search} \label{subsec:nustar_pulsations}
We used \texttt{HENDRICS} to perform the timing analysis. Initially developed as \texttt{MaLTPyNT} \citep{Bachetti2015_MaLTPyNT} for timing analysis of \nustar data, \texttt{HENDRICS} now comprises of tools such as acceleration searches, periodograms, $Z_n^2$ statistics to search for pulsations and extends to some other X-ray missions (e.g., \nicer). We began this analysis by first calibrating the datafile by using the response file for each observation and constructing the light curve using \texttt{HENcalibrate}. The intent here was to check if \target exhibited rapid variability along with modality such that the light curve could be distributed into `high', `low' and `flare' regions as seen in transitional millisecond pulsars. No modality was observed. 

Similar to the techniques used in Appendix~\ref{subsec:nicer_pulsations}, we launched acceleration search using \texttt{PRESTO} to search for periodic pulsations. We split the observation into chunks of 720\,s each and allowed for 5\% overlap within these chunks. We then used \texttt{HENbinary} from \texttt{Hendrics} to render these time series in the format preferred by \texttt{accelsearch}. We binned the light curve to 1\,ms bins. After that, we used the \texttt{accelsearch} routine in \texttt{PRESTO} and searched to a \texttt{zmax} depth of 10 and detection threshold of 2$\sigma$. No viable ``candidates" were detected.

\subsection{Modeling Relativistic Reflection}\label{subsec:relxillCp}
Here we present details of the spectral fitting in Section~\ref{subsec:spectral_joint2}.

% checked
In the \texttt{relxillCp} model, the $\Gamma$ parameter (power law index of the incident spectrum) was fixed at the same value as that in the \texttt{simpcutx} model. 
% checked
The outer disk radius ($R_{\rm out}$) was fixed at a fiducial value of $400\,r_{\rm g}$ \citep{Choudhury2017}, since it has little effect on the X-ray spectrum. Here $r_{\rm g} = GM/c^2$ is the gravitational radius. 
% checked
The electron temperature ($kT_e$) describes the observed high energy cutoff of the spectrum. Since no sign of a power-law cutoff was observed in the \nustar data, $kT_e$ was fixed at the maximum value of 1\,MeV. 
% checked
Redshift ($z$) was fixed at 0 since \target is a Galactic source. 
We included a cross-normalization term (\texttt{constant}) between FPMA, FPMB, and \nicer data. 
% checked
To reduce the complexity of this model, we frozen the reflection fraction ($R_{\rm F}=1$). The inner and outer emissivity index were set at the same value $q$ throughout the accretion disk, making $R_{\rm break}$ obsolete. 

If we fix the black hole spin parameter at $a=0$ or $a=0.998$, and let $R_{\rm in}$, $q$, and $i$ be free, then the fitting will result in parameters loosely constrained, as most of these parameters are correlated \citep{Dauser2013}. Therefore, we experimented by fitting multiple models, and for each model we fixed two of the four parameters. First, we fixed $a=0$, $q=3$, and let $R_{\rm in}$ and $i$ be free. The best-fit values are listed in Table~\ref{tab:august}.

Next, we fixed $a=0.998$, the inclination to the value obtained in the previous fit ($i=27.0^{\circ}$), and allowed $R_{\rm in}$ and $q$ to be free. The best-fit model has similar statistics to that with $a=0$ (Table~\ref{tab:august}). However, this model results in a flatter emissivity law ($q\sim 2.8$) with an inner radius still relatively close to ISCO ($R_{\rm in} \sim 4\pm3 \, R_{\rm ISCO}$). This is contrary to the theoretical expectation of a steep emissivity profile for rapidly rotating black holes with compact coronae, unless the source of power-law photons is placed much farther along the rotational axis, which conversely will result in weaker reflection features (see Fig.~3 in \citealt{Dauser2013}).

Finally, we fixed $a=0$ or $a=0.998$, and $i$ to higher values (45$^\circ$, 60$^\circ$). The fit quality decreases, with clear residuals around the Fe line. Therefore, from the point of view of reflection, the inclination ($i$) of the inner disk is well constrained to $i\lesssim 30^{\circ}$.

\bibliography{19weyx}{}
\bibliographystyle{aasjournal}

\end{document}